\documentclass[useAMS,usenatbib]{mnras}
\usepackage{textcomp,epstopdf}
\usepackage{amssymb,amsmath}
\usepackage{graphicx}
\newcommand{\lcdm}{$\Lambda$CDM}
\newcommand{\Mu}{\mbox{$10^{15}h^{-1}M_\odot$}}
\newcommand{\kcorr}{\hbox{$k$-correction}}
\newcommand{\flux}{\hbox{erg~cm$^{-2}$~s$^{-1}$ }}
\newcommand{\kms}{\hbox{km~s$^{-1}$}}
\newcommand{\XR}{\hbox{X-ray}}
\newcommand{\lx}{\hbox{$L_{\rm X}$}}

\newcommand{\lumin}{\hbox{erg~s$^{-1}$}}
\newcommand{\Om}{\hbox{$\Omega_{\rm m}$}}
\newcommand{\Dvir}{\hbox{$\Delta_{\rm vir}$}}
\newcommand{\Mvir}{\hbox{$M_{\rm vir}$}}
\newcommand{\rvir}{\hbox{$r_{\rm vir}$}}
\newcommand{\cvir}{\hbox{$c_{\rm vir}$}}
\newcommand{\rc}{\hbox{$r_{\rm c}$}}
\newcommand{\rco}{\hbox{$r_{\rm c0}$}}
\newcommand{\ctho}{\hbox{$c_{200\_0}$}}
\newcommand{\Iband}{\hbox{$I$-band}}

\newcommand{\appropto}{\mathrel{\vcenter{\offinterlineskip\halign{\hfil$##$\cr \propto\cr\noalign{\kern2pt}\sim\cr\noalign{\kern-2pt}}}}}
\title[{\bf STRONG LENSING IN THE INNER HALO OF GALAXY CLUSTERS}]
  {Strong Lensing In The Inner Halo Of Galaxy Clusters}
\author[{\bf SAEZ ET AL.}]
{C.~Saez,$^{1,2}$ L.~E.~Campusano,$^3$ E.~S.~Cypriano,$^4$ L.~Sodr\'e,$^4$ and
J.-P.~Kneib.$^{5,6}$ \\
$^1$Korea Astronomy and Space Science Institute (KASI), 61-1, Hwaam-dong, Yuseong-gu, Deajeon 305-348, Republic of Korea \\
$^2$Department of Astronomy, University of Maryland, College Park, MD 20742-2421, USA \\
$^3$Observatorio Astron\'omico Cerro Cal\'an,
Departamento de  Astronom\'ia, Universidad de Chile, Casilla 36-D,
Santiago, Chile \\
$^4$Departamento de Astronomia, IAG , Universidade de
S\~ao Paulo, R. do Mat\~ao 1226, Cidade Universit\'aria,
05508-090, S\~ao Paulo,\\ Brazil \\
$^5$Laboratoire d'Astrophysique, Ecole Polytechnique F\'ed\'erale de Lausanne (EPFL), Observatoire de Sauverny, CH-1290 Versoix,\\ Switzerland\\
$^6$Aix Marseille Universit\'e, CNRS, LAM (Laboratoire d'Astrophysique de Marseille) UMR 7326, 13388, Marseille, France
}

\date{}

\pagerange{\pageref{firstpage}--\pageref{lastpage}} \pubyear{2002}

\def\LaTeX{L\kern-.36em\raise.3ex\hbox{a}\kern-.15em
    T\kern-.1667em\lower.7ex\hbox{E}\kern-.125emX}

\begin{document}

\label{firstpage}

\maketitle

\begin{abstract}
We present an axially symmetric formula to calculate the probability of finding
gravitational arcs in galaxy clusters, being induced by their massive dark matter haloes,
as a function of clusters redshifts and virial masses. The formula includes the ellipticity of the clusters dark matter potential by using a pseudo-elliptical approximation.
The probabilities are calculated and compared for two dark-matter halo profiles,
the Navarro, Frenk and White (NFW) and the Non-Singular-Isothermal-Sphere (NSIS).
We demonstrate the power of our formulation through a Kolmogorov-Smirnov (KS) test on the strong lensing statistics of an X-ray bright sample of low redshift Abell clusters. 
This KS test allows to establish limits on the values of the concentration parameter for the NFW profile ($c_\Delta$) and the core radius for the NSIS profile (\rc), which are related to the lowest cluster redshift ($z_{\rm cut}$) where strong arcs can be observed. 
For NFW dark matter profiles, we infer cluster haloes with concentrations that are consistent to those predicted by $\Lambda$CDM simulations.
As for NSIS dark matter profiles, we find only upper limits for the clusters core radii and thus do not rule out a purely SIS  model.
For alternative mass profiles, our formulation provides constraints through $z_{\rm cut}$ on the parameters that control the concentration of mass in the inner region of the clusters haloes. 
We find that  $z_{\rm cut}$ is expected  to lie in the  0.0--0.2 redshift, highlighting the need to include very low-$z$ clusters in samples to study the clusters mass profiles.
\end{abstract}


\section{Introduction}

Arc statistics is an important tool to test cluster structure
\citep[e.g.,][]{1993MNRAS.262..187W,1995A&A...297....1B,2001ApJ...559..572O,2004ApJ...600L...7H,2015arXiv151104002X}
and cosmology \citep[e.g.,][]{1998A&A...330....1B,2005ApJ...635..795L,2010Sci...329..924J}. As a consequence,
numerous arc surveys have been performed through the use of ground and space
telescopes. These arc searches have been mainly based on X-ray selected clusters  \citep[e.g.,][]{1999A&AS..136..117L,2006EAS....20..269C,2015ApJ...806....4M}, and
clusters chosen in the optical \citep[e.g.,][]{2003ApJ...593...48G,2008AJ....135..664H}. 

In particular, constraining the clusters dark matter halo density profile
has been an objective advocated by many investigations in the
last 15 years. A popular model characterizing their radial mass profile has been
the Navarro, Frenk and White model \citep[NFW;][]{1996ApJ...462..563N}, also known as the ``universal profile''. 
The acceptance of this model is mainly due to its foundation on $\Lambda$CDM $N$-body simulations. The NFW is represented as $\rho(r)=\rho_s(r/r_s)^{-1}(1+r/r_s)^{-2}$, and
therefore, presents central cusps given by $\rho(r)\propto r^{-1}$. Current high-resolution $\Lambda$CDM simulations predict dark matter galaxy clusters haloes with shallow central cusps (${\rm dlog}\,\rho / {\rm dlog}\,r \gtrsim -1.0$). However, as the radius increases, this haloes become progressively steeper and well fitted by a NFW profile \citep[e.g.,][]{2012MNRAS.425.2169G}. 
The baryonic mass is not considered in pure dark matter $\Lambda$CDM simulations. Its presence, could be producing steeper and more concentrated mass profiles in the central regions of the clusters \citep[e.g.,][]{2004ApJ...616...16G}. 
The gravitational effect of the baryonic matter has been recently studied through combined observations of strong lensing, weak lensing and resolved stellar kinematic within the brightest cluster galaxy \citep[BCG;][]{2013ApJ...765...25N,2013ApJ...765...24N}. These studies suggest that although the presence of baryonic dark matter near the cluster centers is significant, pure dark matter models give reliable fits to the total mass distributions. This happens even at scales where baryonic mass should be dominant. 

The modeling of arc and arclets 
also shows discrepant results regarding the mass profile
describing clusters haloes. For example, \cite{2003A&A...403...11G} find that the
cluster MS~2137 is better constrained by an isothermal profile 
 when compared to a NFW model. On the contrary, \cite{2003ApJ...598..804K} show that
the mass distribution of Cl~0024+1654 strongly favors the NFW
profile. Since there is still debate concerning the dark matter halo
profiles of galaxy clusters \citep[e.g.,][]{2013MNRAS.436.2616B}, it is important to develop other
independent methodologies to probe the haloes. Arc statistics
allow the development of 
methodologies that are not sensitive to specific selection effects or 
systematic errors involved in individual observations. 
Besides, it is well known that the study of the lensing
properties of nearby galaxy clusters ($z\lesssim0.5$) offers
an unique opportunity to investigate the cluster central regions with
high spatial resolution \citep[e.g.,][]{1998ApJ...496L..79C,2001AJ....121...10C}.

In this paper, we present an axially symmetric formula \citep[see e.g.,][]{1999ApJ...524..504C, 2000AnP...512..384K, 2001ApJ...559..572O, 2001AJ....121...10C} to calculate the number of arcs produced by dark matter haloes as a function of the cluster's redshift and mass. Our formulation is also used to study departures from axial symmetry by including  a pseudo-elliptical  approximation \citep{2002A&A...390..821G} on the dark energy potential.
Predictions are determined and compared for two currently competing cluster mass models,  the Non-Singular-Isothermal-Sphere (NSIS) and the NFW.
The goal of this work is to find parameters that could be observationally constrained through the redshift distribution of the number of arcs in a low-redshift sample of galaxy clusters.

The layout of the paper is as follows: in \S\ref{S:meto}, we introduce the conventions used and describe our formulation to calculate the number of arcs distribution.  
In \S\ref{S:nar}--\S\ref{S:aarc} we analyze for NSIS and NFW profiles the parameter sensitivity of the number of arcs distribution. 
In \S\ref{S:appl}, using our formulation, we implement a statistical test to provide constraints on parameters of the clusters dark matter haloes. 
This test is applied to a low redshift X-ray bright sample of Abell clusters. 
In \S\ref{S:conc} we summarize our results.
Throughout this paper, unless stated otherwise, we use
cgs units, and we adopt a flat
$\Lambda$-dominated universe with, $\Omega_{\Lambda} = 0.7$,
and $\Omega_{\rm m} = 0.3$.

\section{METHODOLOGY} \label{S:meto}

\subsection{Axially symmetric models in general}\label{S:defl}

The dimensionless lens equation, relating the angular position of the image on the lens plane ($x$) with that of its source on the source plane ($y$) is \citep[see, e.g.,][]{1992grle.book.....S}:
\begin{equation} \label{xy}
y=x-\alpha(x),
\end{equation}
where $\alpha(x)$ is the dimensionless deflection angle, which is given by
\begin{equation} \label{eq:alpha}
\begin{split}
\alpha(x)&=\frac{2}{x}\left(\frac{4\pi
G}{c^2}\frac{D_lD_{ls}}{D_s}\right)\int_0^x
\Sigma(\xi_0x')x'dx'\\
&=\frac{2}{x}\int_0^x \kappa(x') x'dx'.
\end{split}
\end{equation}
Here $\Sigma(\xi)=\int_{-\infty}^{+\infty}\rho(\xi,z)dz$ is the projected surface density, $D_l$, $D_s$ and $D_{ls}$ are the angular distance, to the lens, to the source, and between the lens and the source. Additionally, $\xi_0$ is a length-scale on the lens plane and  $\kappa(x)=\Sigma(\xi_0 x)/\Sigma_{\rm crit}$ ($\Sigma_{\rm crit}=c^2 D_s (4\pi G D_lD_{ls})^{-1}$) is the surface mass density in units of the critical surface density for lensing. Equation (\ref{eq:alpha}) can also be written as $\kappa=(\alpha/x+d\alpha/dx)/2$. 

The amplification factor  of an image will be given by $\mu=|\lambda_t \lambda_r|^{-1}$, where
$\lambda_r$ and $\lambda_t$ are the radial and tangential
eigenvalues of the Jacobian matrix describing the image distortion
in the lens equation. They are expressed as:
\begin{equation}\label{lrt}
\lambda_r=1-\kappa+\gamma \hspace{10pt} {\rm and} \hspace{10pt} \lambda_t=1-\kappa-\gamma,
\end{equation}
with $\gamma=\alpha/x-\kappa$.
Under this formulation, from equations~\ref{xy} and \ref{lrt},
$y(x)=x \lambda_t(x)$. Throughout this paper, we call $x_t$ as the
value of $x$ where $\lambda_t(x_t)=0$ and $x_r$ where
$\lambda_r(x_r)=0$.
For tangentially elongated images
$R_t=\lambda_r/\lambda_t$ is the length-to-width ratio.
The cross section of the tangential image with length-to-width
ratio greater than $R_t$ is given by
\begin{equation} \label{eq:Rtan}
\sigma_{c}(R_t)=2\pi \int_0^{y_{R_t}} y'dy'=\pi y_{R_t}^2.
\end{equation}
Notice that for $x=x_{R_t}$,
$\lambda_r(x_{R_t})/\lambda_t(x_{R_t})=R_t$ and
$y_{R_t}=x_{R_t}\lambda_t(x_{R_t})$. Therefore, equation~\ref{eq:Rtan} can be written as well as
\begin{equation}
\sigma_{c}(R_t)=\pi \lambda_t^2(x_{R_t}) x_{R_t}^2.
\end{equation}
The cross section in the source coordinates is then
\begin{equation}
\hat{\sigma}_{c}=\eta_0^2
\sigma_{c}=\xi_0^2\left(\frac{D_s}{D_l}\right)^2 \sigma_{c},
\end{equation}
where $\eta_0=\xi_0D_s/D_l$ is a length-scale on the source plane.

The condition $\lambda_r(x_r)=0$
($y_r=|\lambda_r(x_r)|$) will limit the region where
strong-lensing is possible. The brightest image in the
strong-lensing region is constrained to $x>x_t$;
$0<y<y_r$, where $x_t$ is given from
$\lambda_t(x_t)\equiv0$ ($y_t=0$).  
The brightest radially elongated image is limited to $x_r<x<x_{r_m}$ and $y_{r_m}<y<y_r$; where $\lambda_t(x_{r_m})/\lambda_r(x_{r_m})=-1$. The cross section for radial images with length-to-width ratio greater than $R_r$ is given by $\sigma_{c}(R_r)=\pi (y_r^2-y_{R_r}^2)$, where $R_r=-{\lambda_t(x_{R_r})}/{\lambda_r(x_{R_r})}$ and $y_{R_r}=y(x_{R_r})$.  Note that the condition  $\mu>1$ limits the region where strong lensing images can form. This is especially important for the formation of radial images.
Throughout this this paper, with exception of Appendix~\ref{S:aprt}, we focus our
analysis of cross sections and the statistics of arcs of the brightest images, which are the tangential. The reason is that in a survey of galaxy clusters the probability of detection of tangential images is approximately an order of magnitude greater than that of radial images (see Appendix~\ref{S:aprt}).

\subsection{Dark matter models}

In this work we adopt two axially symmetric dark matter
mass-models to describe the mass distribution of galaxy clusters:
the Navarro, Frenk and White (NFW) profile, and
the Non-Singular-Isothermal-Sphere (NSIS).  
Here we briefly describe the strong-lensing physical parameters corresponding to these models.
\subsubsection{NFW model} \label{S:NFWm}

The NFW radial and projected mass-density profiles are given by
\begin{equation}
\begin{split}
&\rho(r)=\frac{\rho_s}{(r/r_s)(1+r/r_s)^2} \\
&\Sigma(r_s\, x)=\frac{2 \rho_s
r_s}{1-x^2}\left(\frac{1}{\sqrt{1-x^2}}{\rm
arctanh}\sqrt{1-x^2}-1\right)
\end{split}
\end{equation}
For this profile,  taking $\xi_0=r_s$ and defining $\kappa_s=\rho_s
r_s \Sigma_{\rm crit}^{-1}$ ($ \Sigma_{\rm crit}$ defined in \S\ref{S:defl}), we have that
\[ \kappa(x)=\frac{2\kappa_s}{1-x^2}\left(\frac{1}{\sqrt{1-x^2}}{\rm 
arctanh}\sqrt{1-x^2}-1\right)=\frac{2\kappa_s}{x}\frac{dg}{dx}. \]  
Consequently, the values of the deflection angle, $\gamma$ and eigenvalues are
\begin{equation}
\begin{split}
\alpha(x)&=\frac{4\kappa_s}{x}\left({\rm
ln}\frac{x}{2}+\frac{1}{\sqrt{1-x^2}}{\rm
arctanh}\sqrt{1-x^2}\right)\\ 
&\equiv\frac{4\kappa_s}{x} g, \\
\gamma(x)&=\frac{2\kappa_s}{x}\left(\frac{2g}{x}-\frac{dg}{dx}\right),\\
\lambda_r(x)&=1-\frac{4 \kappa_s}{x}
\left(\frac{dg}{dx}-\frac{g}{x}\right), \\
\lambda_t(x)&=1-4\kappa_s\frac{g}{x^2}.
\end{split}
\end{equation}
For this model  $x_r$, $x_t$, and $x_{R_t}$ are determined numerically.
Additionally,  there are no restrictions on $\kappa_s$ in order
to produce strong-lensing, albeit $y(x_r)\rightarrow0$ as
$\kappa_s\rightarrow0$. Therefore, given
the scale size of a halo model, there will be a value $\kappa_{\rm smin}$
such that if  $\kappa_s \lesssim \kappa_{\rm smin}$ there will not be  strong lensing effects. For the calculations performed in this  work a value of $\kappa_{\rm smin}=0.08$ has been adopted.

\subsubsection{SIS and NSIS models} \label{S:NSISm}

The non-singular-isothermal-sphere or NSIS model is a
generalization of the SIS model, with the addition of a core
radius to avoid the density singularity in the origin. The NSIS
profile is given by
\begin{equation}
\rho(r)=\frac{\sigma_v^2}{2\pi G (r^2+r_c^2)}, \hspace{20pt}
\Sigma(\xi)=\frac{\sigma_v^2}{2 G \sqrt{\xi^2+r_c^2}}.
\end{equation}
Choosing $\xi_0=4 \pi (\sigma_v/c)^2D_lD_{ls}/D_s$ as the
length-scale\footnote{For this choice of length scale $\theta_0=\xi_0/D_l=4 \pi (\sigma_v/c)^2D_{ls}/D_s$ corresponds to the Einstein radius for the SIS model ($r_c=0$).}, $\kappa(x)=1/(2\sqrt{x^2+x_c^2})$; therefore the
deflection angle, $\gamma$, and eigenvalues are:
\begin{equation}
 \label{NFWterms}
\begin{aligned}
&\alpha(x)=\frac{\sqrt{x^2+x_c^2}-x_c}{x},\\
&\gamma(x)=\frac{x^2+2x_c(x_c-\sqrt{x^2+x_c^2})}{2x^2 \sqrt{x^2+x_c^2}} \\
&\lambda_r(x)=1+\frac{x_c^2-x_c\sqrt{x^2+x_c^2}}{x^2
\sqrt{x^2+x_c^2}},\\
&\lambda_t(x)=1-\frac{\sqrt{x^2+x_c^2}-x_c}{x^2}.
\end{aligned}
\end{equation}
The condition of strong lensing is produced when $x_{\rm c}<1/2$. For this model, $x_r=(x_c-(x_c^2+x_c^{3/2}\sqrt{4+x_c})/2)^{1/2}$ and 
$x_t=\sqrt{1-2x_c}$. In
the regime where $x_{\rm c}<1/2$ the area of strong-lensing
corresponds to $-y_r<y<y_r$; additionally $x_{R_t}$ is obtained from
\begin{equation}
\kappa(x_{R_t})\!=\!\frac{Q_t\!+\!1\!-\!2x_c\!-\!\sqrt{(Q_t\!+\!1\!-\!2x_c)^2\!-\!8(Q_t\!-\!1)x_c}}{4(Q_t\!-\!1)x_c}, 
\end{equation}
with
\[ Q_t=(R_t+1)/(R_t-1) \quad{\rm and}\quad \kappa(x_{R_t})=1/\left(2\sqrt{x_{R_t}^2+x_c^2}\right).\]
In the SIS (NSIS
model with $r_c=0$) case $x_r=0$, $x_t=1$,
$\alpha(x)=x/|x|$, $\lambda_r(x)=1$,
$\lambda_t(x)=1-1/|x|$, and the condition for strong
lensing (multiple images) is in the region $-1<y<1$. The
parameterization of the images is $y=x-1$ with magnification
$\mu(x)=|x|/||x|-1|$. For the tangential image
$x~\epsilon~[1,2]$, $x_{R_t}=R_t/(R_t-1)$ and $y_{R_t}=1/(R_t-1)$.

\subsection{Cluster Mass} \label{S:Mass}

The enclosed mass within radius $r$ ($M(r)=4\pi
\int_0^rr^2\rho(r)dr$) for each density profile is
\begin{subequations}
\begin{align}
&M_{\rm NFW}(r)=4 \pi \rho_sr_s^3 \left(-\frac{x}{1+x}+{\rm
ln}(1+x)\right), \: x=\frac{r}{r_s} \\
&M_{\rm NSIS}(r)=\frac{2 \sigma_v^2 r_c}{G}\left(x-{\rm arctan}
x \right), \: x=\frac{r}{r_c}.
\end{align}
\end{subequations}
As convention, we define the mass $M_\Delta$  as that encircled by galaxy
cluster when it reaches a radius where its density is a factor
$\Delta$ of the critical
density $\rho_{\rm crit}$,  which is given by
\begin{subequations}
\begin{align}
&\rho_{\rm crit}=\frac{3H^2(z)}{8\pi G}=\rho_{\rm crit,0}E^2(z) \\
&\rho_{\rm crit,0}=\frac{3H_0^2}{8\pi G}\approx1.88\cdot10^{-29}h^2 ~
{\rm g~cm}^{-3} \\
&
E^2(z)=\Omega_{\rm m}(1+z)^3+(1-\Omega_{\rm m}-\Omega_{\Lambda})(1+z)^2+\Omega_{\Lambda}.
\end{align}
\end{subequations}
We use two popular choices for the overdensity factor $\Delta=200$ and $\Delta=\Delta_{\rm vir}$.
The virialized  overdensity factor is estimated based on the assumption of spherical collapse by $\Delta_{\rm vir}(z)=178\,\Om(z)^{0.45}$ \citep[e.g.,][]{1998ApJ...503..569E,2000ApJ...538..477N,2001ApJ...554..114E}.
In this last expression, $\Omega_{\rm m}(z)$ is the matter cosmological parameter as a function of
redshift and is given by
\begin{equation}
\Om(z)=\frac{\Omega_{\rm m}\,(1+z)^3}{E^2(z)} \hspace{10pt}.
\end{equation}
Defining $r_\Delta$ as:
 \begin{equation} \label{eq:rvir}
r_\Delta=\left(\frac{3 M_\Delta}{4\pi \Delta \rho_{\rm crit}}\right)^{1/3}, 
\end{equation}
we can find relations between an
estimate of the cluster mass  such as $\Mvir$$\equiv$$M(\rvir)$ or $M_{200}$$\equiv$$M(r_{200})$, 
and the parameters that define each of the mass-density profiles.

In the case of the NFW model, choosing
$\rho_s=\rho_{\rm crit}\delta_{\rm c}$,
\[ r_s=\frac{r_\Delta}{c_\Delta}\approx \frac{9.51}{c_\Delta} \frac{M_{\Delta 15}^{1/3}}{\Delta^{1/3} E(z)^{2/3}}h^{-1}\, {\rm Mpc}, \] 
with $M_{\Delta 15}=M_\Delta/M_{15}$ ($M_{15}=10^{15} h^{-1} M_\odot$),
and using the definition of $r_\Delta$, the
following relation holds for the parameters $c_\Delta$ and
$\delta_c$ \citep{1996ApJ...462..563N}:
\begin{equation}
\delta_c=\frac{\Delta \, c_\Delta^3}{3\,[{\rm ln}(1+c_\Delta)-c_\Delta/(1+c_\Delta)]}. \\
\end{equation}
Notice that $c_\Delta$ and $\delta_c$ are dimensionless parameters
and $c_{\Delta}$ is usually referred as the concentration
parameter.\footnote{Originally defined by \cite{1996ApJ...462..563N} for
$\Delta$=200.} In general $c_\Delta$  can be expressed as:
\begin{equation} \label{eq:ctho}
c_{\Delta}(M_\Delta,z) = c_{\Delta 0} M_{\Delta 15}^{\alpha_M} (1+z)^{\alpha_z},  
\end{equation}
where $c_{\Delta 0}=c_{\Delta}(M_{15},0)$. Throughout this paper, we will use $\cvir_0$ to identify $c_{\Delta 0}$ when $\Delta=\Delta_{\rm vir}$ and $c_{200\_0}$ to recognize $c_{\Delta 0}$ when $\Delta=200$.
Additionally, unless stated differently, we estimate $c_\Delta$ from the \lcdm\ $N$-body simulations \cite{2008MNRAS.390L..64D} in the case of relaxed clusters.  
In these estimations $\alpha_M\approx -0.09$ (independent of $\Delta$), $\alpha_z\approx-0.7$ for $\Delta=\Delta_{\rm vir}$, $\alpha_z\approx -0.4$ for $\Delta=200$, $\cvir_0 \approx 5.3$, and $c_{200\_0} \approx 3.8$.
Observations on the X-ray proprieties of virialized clusters are consistent with  the evolution
of $c_\Delta$ \citep[e.g.,][]{2007MNRAS.379..209S}  , however, these observations in general predict steeper dependencies of $c_\Delta$ with $M_\Delta$. For example from the works of \cite{2007ApJ...664..123B, 2007MNRAS.379..209S} and \cite{2010A&A...524A..68E}, $c_\Delta\appropto\ M_\Delta^{\alpha_M}$  with $\alpha_M \lesssim -0.2$.

For the NSIS model, assuming that $r_\Delta\gg r_c$, the
velocity dispersion of the cluster is
\begin{equation} \label{eq:sv1}
\begin{aligned}
\sigma_v&=f_{\sigma}\sqrt{\frac{M_\Delta G}{2r_\Delta}}\\
&=476 f_\sigma \left(\frac{M_\Delta}{M_{15}}\right)^{1/3} [\Delta(z) E^2(z)]^{1/6}~{\rm km~s^{-1}}.
\end{aligned}
\end{equation}
Simulations in general do not predict isothermal sphere models;
therefore to obtain $\sigma_v$ we have introduced a factor
$f_\sigma$, which we choose equal to 0.78 
(similar to \cite{1998ApJ...495...80B}). Through work, for NSIS models we estimate $r_c$ as proportional to the virial radius of the cluster (see equation~\ref{eq:rvir}), i.e.,
\begin{equation} \label{eq:rcmz}
r_{\rm c}(M_\Delta, z)=r_{\rm c0}\left(\frac{\Mvir(M_\Delta)}{M_{15}}\right)^{\frac{1}{3}}\left(\frac{E(0)}{E(z)}\right)^{\frac{2}{3}}\left(\frac{\Dvir(0)}{\Dvir(z)}\right)^{\frac{1}{3}},
 \end{equation}
where $\Mvir(M_\Delta)=(\Delta/\Delta_{\rm vir})^{1/2}M_\Delta$ and $r_{\rm c0}$ is the core radius at $z=0$ for $\Mvir=M_{15}$.

\subsection{Strong lensing arcs statistics} \label{Nasc}

The total number of arcs produced by a lens (galaxy cluster) with redshift $z_L$ depends on numerous parameters. 
The most important are: parameters from the dark matter profile, the comoving density of the galaxies at $z_s>z_L$, and the brightness detection limit (size of the telescope).
Assuming $n_0(\bar{\mu},z_s)$ as the comoving density of galaxies at redshift
$z_s$, and $\bar{\mu}$ as the image brightness amplification, the total number of detected arcs is obtained by:
\begin{equation} 
N_{\rm arcs}(M,z_L)\!=\!\int_{z_L}^{z_{\rm max}}\!n_o(\bar{\mu},z_s)\hat{\sigma}_c(M,z_L,z_s)\frac{cdt}{dz_s}(1\!+\!z_s)^3dz_s. \label{eq:naog}
\end{equation}
As in \cite{2001ApJ...559..572O} we assume that $\bar{\mu}=R_t \lambda^{-2}_r(x_t)$ for tangentially elongated images  and $\bar{\mu}=R_r \lambda^{-2}_t(x_r)$ for radially elongated images (see Appendix~\ref{S:aprt}). 
In equation~\ref{eq:naog}  $z_{\rm max}$ is
the maximum redshift assumed for the galaxies, and $cdt/dz_s$
denotes the proper length differential at $z_s$, i.e.
\begin{equation}
\frac{cdt}{dz_s}=\frac{c}{H(z_s)(1+z_s)}=\frac{c}{H_0E(z_s)(1+z_s)}.
\end{equation}
Additionally $n_0(\bar{\mu},z)$ is estimated by
\begin{equation} \label{no1}
n_0(\bar{\mu},z_s)=\int_{\frac{L_{\rm min}}{\bar{\mu}}}^{\infty}\phi(L,z_s)dL,
\end{equation}
where $L_{\rm min}(z)$ is the minimum luminosity of a galaxy
in order to be detected as an image. Additionally,
$\phi(L,z)$ is the comoving density of galaxies
which is represented  by a Schechter
function \citep{1976ApJ...203..297S}, i.e.:
\begin{equation} \label{phi}
\phi(L,z)=\phi^*(z_s)\left(\frac{L}{L^*(z_s)}\right)^{\alpha(z_s)}{\rm
exp}\left(\frac{L}{L^*(z_s)}\right) \frac{dL}{L^*(z)}
\end{equation}
Therefore, using equation~\ref{phi} to integrate equation~\ref{no1} we obtain
\begin{equation} \label{no2}
n_0(\bar{\mu},z_s)=\phi^*(z_s)
\Gamma\left[1+\alpha(z_s),\bar{\mu}^{-1}\frac{L_{\rm min}(z_s)}{
L^*(z_s)}\right],
\end{equation}
where $\Gamma$ is the incomplete gamma function. In order to
express  $L_{\rm min}(z_s)/L^*(z_s)$ in magnitudes we use
\begin{subequations} \label{mags}
\begin{align}
M^*(z)&=-2.5\,{\rm log}\frac{L^*(z)}{\rm 4\pi(10pc)^2}+{\rm const},{\rm \:and}  \label{magsa} \\
m_{\rm lim}&=-2.5\,{\rm log}\frac{L_{\rm min}(z)}{4\pi
D_s^2(1+z_s)^4}+k(z)+{\rm const}, \label{magsb}
\end{align}
\end{subequations}
where $k(z)$ is the \kcorr\ in a given band-pass. Finally, combining
equations~\ref{magsa} and \ref{magsb} we get
\begin{equation} \label{no3}
\begin{split}
{\rm log}\left( \frac{L_{\rm min}(z_s)}{
L^*(z_s)}\right)\!=\!-\frac{2}{5}
\{&m_{\rm lim}\!-\!5{\rm log}\left[D_s(1\!+\!z_s)^2\right]\\
&\!-\!25-M^*(z_s)\!-\!k(z_s) \},
\end{split}
\end{equation}
where $D_s$ is in units of Mpc.

\subsection{Pseudo-elliptical modeling} \label{Pseudo}

In order to estimate the effect of ellipticity in the lens statistics, we partially depart from  axially symmetric models using a pseudo elliptical approach \citep{2002A&A...390..821G}. In order to estimate strong lensing parameters we use:
\begin{subequations}
\begin{align}
\kappa_{\epsilon}(\bf x)=&\kappa({\bf x_{\epsilon}})+\epsilon~{\rm cos2\phi_{\epsilon}}~\gamma({\bf x_\epsilon}) \label{eq:kape} \\ 
\begin{split}
\gamma_\epsilon^2(\bf x)=&\gamma^2({\bf x_\epsilon})+2\epsilon~{\rm cos} 2\phi_\epsilon \gamma({\bf x_\epsilon}) \kappa({\bf x_\epsilon}) \\
&+\epsilon^2 [\kappa^2({\bf x_\epsilon})-{\rm sin}^22\phi_\epsilon\gamma^2({\bf x_\epsilon})], \label{eq:game}
\end{split}
\end{align}
\end{subequations}
where $\kappa_\epsilon$ and $\gamma_\epsilon$ are $\kappa$ and $\gamma$ transformed from the spherically symmetric cases \citep[see e.g.,][]{2012A&A...544A..83D}.
These expressions are obtained assuming that the elliptical surface mass distribution depends on ${\bf x_\epsilon}=\sqrt{x_{1\epsilon}^2+x_{2\epsilon}^2}$, with $x_{1 \epsilon}=\sqrt{a_{1\epsilon}}x_1$, $x_{2 \epsilon}=\sqrt{a_{2\epsilon}}x_2$, $\phi_\epsilon={\rm atan}(x_{2\epsilon}/x_{1\epsilon})$, $a_{1 \epsilon}=1-\epsilon$, and $a_{2 \epsilon}=1+\epsilon$.
The ellipticity $\epsilon$ of the pseudo-elliptical model differs from the standard ellipticity $\epsilon_\Sigma$  expected from a purely elliptical model.
 From equations~\ref{lrt}, \ref{eq:kape} and \ref{eq:game} we estimate the parameters $\lambda_t$ and $\lambda_r$  to  calculate cross sections in the image plane. Additionally, we transform these curves to the source plane by using the following transformation:
\begin{equation}
\left(
    \begin{array}{c}
      y_1 \\
      y_2
    \end{array}
  \right) = 
  \left(
  \begin{array}{c}
      x_1 \\
      x_2
    \end{array}
     \right) + \alpha({\bf x_\epsilon}) 
     \left(
    \begin{array}{c}
      \sqrt{a_{1\epsilon}}{\rm cos}\phi_\epsilon \\
     \sqrt{a_{2\epsilon}}{\rm sin}\phi_\epsilon
    \end{array}
  \right),
  \end{equation}
where $\alpha(x)$ is obtained from equation~\ref{eq:alpha}. 
Through this process we compute $\hat{\sigma}_c$ and $\bar{\mu}$ in the source plane to estimate the statistics of arcs by using equation~\ref{eq:naog}.  For the pseudo elliptical case  $\bar{\mu}$ is averaged as a function of its polar angle in the source plane i.e.,
\begin{equation}
\bar{\mu}=\frac{2R_t}{\pi} \int_0^{\pi/2}\lambda^{-2}_r(y_t(\phi))d\phi, 
\end{equation}
where $y_t$ is the tangential critical curve in the source plane.

The pseudo elliptical model is a good approximation of an elliptical potential for small values of $\epsilon$. However, at high values of $\epsilon$ the mass density profiles tend to become peanut shaped \citep[as noticed earlier by][]{2002A&A...390..821G}.
Using similar analysis as \cite{2012A&A...544A..83D}, we find that the pseudo elliptical approximation is valid for $\epsilon \lesssim 0.3$ ($\epsilon_\Sigma \lesssim 0.5$) for both the NSIS and NFW models (see Appendix~\ref{S:apps}). 
Additionally, we confirm (as in \citealp{2012A&A...544A..83D} and \citealp{2002A&A...390..821G}) that albeit $\epsilon$ is similar to $\epsilon_\Sigma$ for low values of $\epsilon$, $\epsilon$ is smaller than $\epsilon_\Sigma$ in the ranges of ellipticities where the pseudo-elliptical approximation is valid (see Appendix~\ref{S:apps}).
\begin{figure}
   \includegraphics[width=8.4cm]{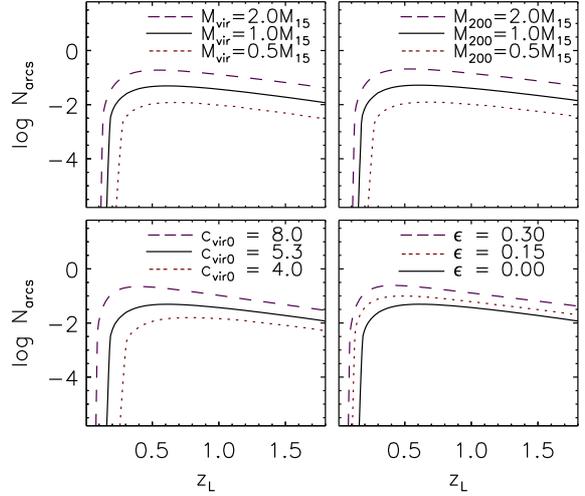}
        \centering
       \caption{Logarithm of number of arcs versus lens redshift in the NFW model.
                The upper left (right) corresponds to 
                tangential arcs of galaxy clusters with three different \Mvir\ 
               ($M_{200}$) masses: $5\!\cdot\!10^{14}h^{-1}M_\odot$ 
              (dotted line),  $10^{15} h^{-1}  M_\odot$ 
            (solid line) and $2\!\cdot\!10^{15}h^{-1}M_\odot$ (dashed line). 
                For the lower left panel we generate curves for 
                for $\cvir_0=4.0$ 
                (dotted line), 5.3 (solid line) and 8.0 (dashed line). 
              For the lower right panel we generate curves for $\epsilon=0.00$ 
              (solid line), 0.15 (dotted line) and 0.30 (dashed line).
                 In the two upper and the lower right panels \cvir\ and 
             $c_{200}$ are from \protect\cite{2008MNRAS.390L..64D}, 
               in the lower panels $\Mvir=10^{15} h^{-1}M_\odot$,  
                 and in the lower left panel  $\cvir\propto(1+z)^{-0.7}$. 
              In all insets $m_{\rm lim}=24$ and $R_t=10$.}
     \label{fig:NFWa}
     \end{figure}
\begin{figure}
   \includegraphics[width=8.4cm]{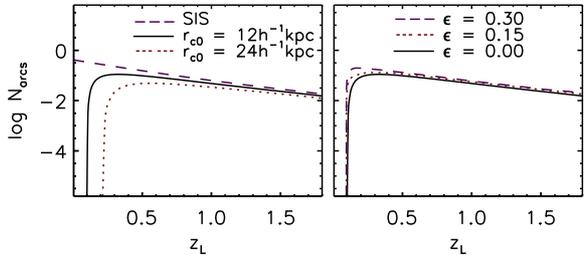}
        \centering
       \caption{Logarithm of number of arcs model versus lens redshift for the NSIS model.
       The left panel corresponds to
       tangential arcs with $\rco=0$ (SIS model; dashed line), $12\,h^{-1}{\rm kpc}$ (solid line) and $24\,h^{-1}{\rm kpc}$ (dotted line).
       The right panel corresponds to
       tangential arcs of haloes with $\rco=12\,h^{-1}{\rm kpc}$ and $\epsilon=0.00$ (solid line), 0.15 (dotted line) and 0.30 (dashed line).  In all insets $M_\Delta=10^{15}\,h^{-1}M_\odot $, $m_{\rm lim}=24$, $R_t=10$, and the dependency $r_c$ with $z_L$ is given by equation~\ref{eq:rcmz}.}
     \label{fig:NSISa}
     \end{figure}

\begin{table}

\begin{center}
\caption{Parameters of the luminosity function used in this work.\label{tab:lumf}}
\begin{tabular}{cccccccccccccc} 
\hline\hline

{\sc Ref.}$^{\rm a}$ & {\sc Redshift} & {\sc BP} &  $M_{\rm I}^*\!-\!5\,{\rm log}h$$^{\rm b}$ &
$\phi^*\,^{\rm c}$ &  $\alpha$ \\

\hline

1 & 0.00--0.45 & $i'$ &  $-21.94$ &  12.8 & $-1.25$\\
2 & 0.45--0.81 & $g'$ & $-22.23$ &  12.2 & $-1.25$\\
2 & 0.81--1.11 & $B$ & $-22.67$ &  11.7 & $-1.25$ \\
2 & 1.11--1.61 & $B$ & $-23.01$ &  7.0 & $-1.25$ \\
2 & 1.61--2.15 & $u'$ & $-22.62$ &  9.6 & $-1.07$ \\
2 & 2.15--2.91 & $u'$ & $-23.42$ & 9.3 & $-1.07$ \\
2 & 2.91--4.01 & $u'$ & $-23.49$ &  6.4 & $-1.07$ \\
2 & 4.01--5.01 & $u'$ & $-23.84$ &  2.3 & $-1.07$ \\

\hline
\end{tabular}
\end{center}

$^{\rm a}$Reference of the luminsity function used: 1$\equiv$\cite{2001AJ....121.2358B} and 2$\equiv$\cite{2004A&A...421...41G}.\\
$^{\rm b}$Obtained from the luminosity functions indicated in the  first column, transforming $M_{\rm BP}^*$ using the following filter transformations: $i'-I=0.68$; $g'-I=1.53$; $B-I=1.88$ and $u'-I=2.69$ \citep[valid for a Sbc galaxy;][]{1995PASP..107..945F}. \\
$^{\rm c}$Units of $10^{-3}h^{3}\rm Mpc^{-3}$.

\end{table}


\section{Results}\label{resu}

Using the assumptions of \S\ref{S:meto} in \S\ref{S:nar} we obtain
results of the estimated strong-lensing  number of arcs
distribution. Additionally, to  clarify the dependence of the arcs statistics
 with  $\hat{\sigma}_c$ (see equation~\ref{eq:naog}), in \S\ref{cro}
we perform a  cross sections analysis in \S\ref{cro}.  
For the results presented in this section we consider the
\Iband. We base the calculation of equation~\ref{eq:naog} on the
luminosity functions of \cite{2001AJ....121.2358B} ($z<0.45$) and \cite{2004A&A...421...41G}
($z>0.45$). As it can be seen from Table~\ref{tab:lumf}, the
luminosity function selected is in shorter comoving wavelength
band-pass (BP from $i'$ for $z=0.00\!-\!0.45$ to $u'$ for $z=4.01\!-\!5.01$) as
the redshift increases. This is in order to approximately follow
the comoving luminosity function of the galaxies that will be
detected in the observer-frame \Iband. 
We have performed transformations of the band
pass magnitudes taking the expressions found in \cite{1995PASP..107..945F} and using
a Sbc as the ``standard galaxy''. The assumed \kcorr\ is from
\cite{1997A&AS..122..399P}; based on this work we have used the following
polynomial approximation to obtain the \kcorr\ of a Sbc
galaxy
\[ k(z_s)=0.01-0.23z_s+1.40z_s^2-0.50z_s^3+0.05z_s^4. \] For the
calculation of equation~\ref{eq:naog} we adopt  $z_{\rm max}=6$ and from hereafter
unless stated differently we assume $R_t=10$ and $m_{\rm lim}=24$. 
Additionally, based on the analysis of
\cite{2001ApJ...559..572O}, the calculations performed in this section 
have been corrected for the fact that the background
lensed galaxies have a finite size (of radius $r_{\rm gal}$). 
This is performed by forcing $\hat{\sigma}_c(M,z_L,z_s)=0$ when $z_s$ is such that 
$\hat{\sigma}_c(M,z_L,z_s)<\pi\,r_{\rm gal}^2$, assuming 
$r_{\rm gal}=1\,h^{-1}{\rm kpc}$.\footnote{It is not expected to find strong lensing effects when the source is bigger or comparable to the strong lensing cross section \citep[e.g.,][]{1992grle.book.....S}.} The chosen value of $r_{\rm gal}$ is close to the size 
of the smallest galaxy observed as an arc \citep{1997A&A...319..764H}. 
\begin{table*}
\begin{minipage}{175mm}
\begin{center}
\caption{$N_{\rm arcs}(z_L)$ curve properties in the NFW models.\label{tab:zNFW}}

\begin{tabular}{ccccccccc}
\hline\hline
\Mvir$^{\rm a}$ & $M_{200}$$^{\rm a}$ & 
$\cvir_0$$^{\rm b}$  & $c_{200\_0}$$^{\rm b}$  & $\epsilon$  & 
 $\epsilon_g$ & $z_{\rm cut}$$^{\rm c}$
&  $z_{\rm peak}$$^{\rm c}$   &  
{\sc reference for} $c_{\Delta}(z)$ \\
\hline

0.5 & ... &  5.3 & ... & 0.00 & 0.00 & 0.13 &  0.68  &   \cite{2008MNRAS.390L..64D}   \\
1.0 & ... &  5.3 & ... & 0.00 & 0.00 & 0.10 &  0.60   &    \cite{2008MNRAS.390L..64D}   \\
2.0 & ... &  5.3 & ... & 0.00 & 0.00 & 0.08  &  0.55  &  \cite{2008MNRAS.390L..64D}  \\
... & 0.5 & ... & 3.8 & 0.00 & 0.00 & 0.13 &  0.69  &   \cite{2008MNRAS.390L..64D}   \\
... & 1.0 & ... & 3.8 &  0.00 & 0.00 & 0.10 &  0.61   &    \cite{2008MNRAS.390L..64D}  \\
... & 2.0 & ... & 3.8 & 0.00 & 0.00 & 0.07  &  0.53  &  \cite{2008MNRAS.390L..64D}  \\

1.0 & ... & 4.0 &  ... & 0.00 & 0.00 & 0.16 & 0.78 &   $\propto$$(1+z)^{-0.7}$ \\
1.0 & ... & 5.3 & ... & 0.00 & 0.00 & 0.10 & 0.61 &  $\propto$$(1+z)^{-0.7}$ \\
1.0 & ... & 8.0 & ... &  0.00 & 0.00 & 0.05 & 0.41 &  $\propto$$(1+z)^{-0.7}$\\

1.0 & ... &  5.3 & ... & 0.15 & 0.00 & 0.07 &  0.50   &    \cite{2008MNRAS.390L..64D}   \\
1.0 & ... &  5.3 & ... & 0.30 & 0.00 & 0.06 &  0.42  &  \cite{2008MNRAS.390L..64D}  \\

1.0 & ... &  5.3 & ... & 0.00 & 0.25 & 0.10 &  0.60   &    \cite{2008MNRAS.390L..64D}   \\
1.0 & ... &  5.3 & ... & 0.00 & 0.50 & 0.10 &  0.60   &    \cite{2008MNRAS.390L..64D}   \\

\hline
\end{tabular}
\end{center}

In these models \hbox{$m_{\rm lim}\!=\!24$} and \hbox{$R\!=\!10$}.\\  
$^{\rm a}$\Mvir\ and $M_{200}$ are given in units of \Mu.\\
$^{\rm b}$ $\cvir_0$ and \ctho\ are defined in \S \ref{S:Mass}.\\
$^{\rm c}$ $z_{\rm cut}$ and $z_{\rm peak}$ are obtained
from the curves presented in Figure~\ref{fig:NFWa}.\\
\end{minipage}
\end{table*}

\begin{table}
\begin{center}
\caption{$N_{\rm arcs}(z_L)$ curve properties in the NSIS models.
\label{tab:zNSIS}}

\begin{tabular}{ccccccc}
\hline\hline

$r_{{\rm c}0}$$^{\rm a}$ & $\epsilon$ &  $\epsilon_g$ &  $z_{\rm cut}$$^{\rm b}$  &
$z_{\rm peak}$$^{\rm b}$ \\
\hline

0.0 & 0.00 & 0.00 & 0.00 & 0.00  \\
12.0 & 0.00 & 0.00 & 0.10 & 0.33  \\
24.0 & 0.00 & 0.00 & 0.20 & 0.57 \\
12.0 & 0.15 & 0.00 & 0.10 &  0.26\\
12.0 & 0.30 & 0.00 &  0.10 &  0.17 \\
12.0 & 0.00 & 0.25 & 0.10 & 0.33  \\
12.0 & 0.00 & 0.50 & 0.10 & 0.33  \\
\hline
\end{tabular}
\end{center}

For these models:
\hbox{$m_{\rm lim}\!=\!24$}, \hbox{$R\!=\!10$} and 
\hbox{$\Mvir \!=\!\Mu$}.\\ 
$^{\rm a}$In units of $h^{-1}\,{\rm kpc}$.\\ 
$^{\rm b}$$z_{\rm cut}$ and $z_{\rm peak}$ are obtained
from the curves presented in Figure~\ref{fig:NSISa}. 

\end{table}


\subsection{Number of arcs distribution.}\label{S:nar}

Our goal is  to study the distribution of $N_{\rm arcs}(z_L)$ (i.e. equation~\ref{eq:naog}) for a relatively local sample of galaxy clusters ($z_L<0.3$). 
Therefore, we are interested in finding parameters of the dark matter profiles that make an impact on $N_{\rm arcs}$  at relatively low redshifts. An important property of $N_{\rm arcs}(z_L)$ is the minimum redshift of the lens (galaxy cluster) at which we can detect strong lensing arcs ($z_{\rm cut}$).
$z_{\rm cut}$ depends mainly on the mass and on halo parameters that modify the concentration of mass of the cluster near its center.  
The approximate parametrical dependencies of $z_{\rm cut}$ can be obtained by assuming $z_{\rm cut} \ll 1$,  $z_{\rm cut} \ll z_s$, and the limiting value of the parameter that imposes the strong lensing condition. Consequently, for the NFW profile assuming  $\kappa_s=\kappa_{\rm smin}$ (see \S~\ref{S:NFWm}) we obtain
\begin{equation} \label{eq:zcNFW}
\begin{aligned}
z_{\rm cut} &\approx\frac{\kappa_{\rm smin}}{3^{1/3}(4\pi)^{2/3}} \left( \frac{cH_0}{G\,\rho_{\rm crit}^{2/3} } \right)  \frac{\Delta_0^{1/3} c_\Delta(M_\Delta,0)}{  M_\Delta^{1/3} \delta_{c0}} \\
           &\approx 16.8\, \Delta_0^{1/3} M_{\Delta15}^{\alpha_M-1/3} \delta_{c0}^{-1} c_{\Delta0},
\end{aligned}
\end{equation}
where $\Delta_0=\Delta(z=0)$, $\delta_{c0}=\delta_c(z=0)$ and $c_\Delta$ is obtained through equation~\ref{eq:ctho}. Additionally, the expression $\delta_{c0}^{-1} c_{\Delta 0}$ is monotonically decreasing function of $c_{\Delta 0}$.
For the NSIS profile assuming  $x_c=1/2$ (see \S~\ref{S:NSISm}) we obtain
\begin{equation} \label{eq:zcNSIS}
\begin{aligned}
z_{\rm cut}&\approx\frac{3^{1/3}}{4^{1/3}\pi^{4/3}}\left( \frac{cH_0}{Gf_\sigma^2 \rho_{\rm crit}^{1/3}} \right)  \frac{r_c(M_\Delta,0)}{\Delta_0^{1/3} M_\Delta^{2/3}} \\
   &\approx 0.0162 \,\Delta_0^{-1/6}  M_{\Delta15}^{-1/3} \left( \frac{r_{c0}}{h^{-1}{\rm kpc}} \right),
\end{aligned}
\end{equation}
where in the last step we have used equation~\ref{eq:rcmz}.

For a fixed mass, variations of the NFW parameter $c_\Delta$ and the NSIS parameter $r_c$ change the fraction of mass encircled at a fixed radius near the center of the cluster. 
For low redshift ($z_L \lesssim 0.2$) NFW lenses with $M_{\rm vir}=M_{15}$ and  $\cvir_0$ equal to 3 and 8,  $M(0.01\,\rvir)/\Mvir$ is $\approx0.0007$ and $\approx0.0022$ respectively. Additionally, $z_{\rm cut}$ are 0.12 and 0.03 respectively.
The SIS profile (i.e. $r_c=0$) corresponds to a NSIS profile with maximal encircled mass near the halo center ($r  \ll \rvir$) with $z_{\rm cut}=0$.  
For NSIS ($z_L \lesssim 0.2$) lenses with mass equal to $M_{15}$ and values of \rco\  equal  0 and $24h^{-1}{\rm kpc}$ ($r_{\rm c}\approx0.012\,\rvir$),  $M(0.01\rvir)/\Mvir$  are $\approx 0.0100$  and $\approx0.0017$ respectively. Additionally, $z_{\rm cut}$ are 0.00 and 0.20 respectively. 
In general, changes in parameters that concentrate in the vecinities of the halo center (at $r \ll \rvir$) should shift $z_{\rm cut}$ to lower values. This is expected given that the strong lensing regime in low redshift clusters is probing their dark matter haloes in the inner regions.\footnote{As an example lets take a simple SIS model. For this case the strong lensing effects will happen inside the Einstein radius which will be at $r_E= 4\pi(\sigma_v/c)^2 D_lD_{ls}/D_s$. If we assume $\Mvir=M_{15}$ (or $\sigma_v \sim 1000\kms$) and $z\lesssim 0.2$ we obtain $r_E\approx400 z h^{-1} {\rm kpc}$, therefore $r_E/\rvir \lesssim 0.2 z$ for $z\lesssim 0.2$.}

In Figure \ref{fig:NFWa}, we present the number of arcs in the NFW
model versus lens redshift. For this Figure, the lower-right panel are cases with ellipticity and the rest of the panels correspond to axially symmetric profiles.  The upper left panel corresponds to arcs for dark matter profiles with three different
virial masses; $\Mvir=5\!\cdot\!10^{14}h^{-1}M_\odot$, $10^{15} h^{-1}
M_\odot$ and $2\!\cdot\!10^{15}h^{-1}M_\odot$. The upper right panel corresponds to arcs for dark matter profiles with three different 
$M_{200}$ masses; $M_{200}=5\!\cdot\!10^{14}h^{-1}M_\odot$, $10^{15} h^{-1}
M_\odot$ and $2\!\cdot\!10^{15}h^{-1}M_\odot$. The lower left panel corresponds to dark matter haloes
with  $M_{\rm vir}=10^{15}h^{-1}M_\odot$, and three values of the concentration parameter at $z=0$
($\cvir_0=4.0$, 5.3 and 8). The lower right panel corresponds to dark matter haloes
with  $M_{\rm vir}=10^{15}h^{-1}M_\odot$, and three values of the ellipticity parameter at $\epsilon=0.00$, 0.15 and 0.30.
In the two upper and the lower right panels we have used
$\cvir$ and $c_{200}$ from \cite{2008MNRAS.390L..64D}, and in the lower left panel of
Figure~\ref{fig:NFWa} we have assumed $\cvir (z)= \cvir_0 (1+z)^{-0.7}$.

As seen from  Table~\ref{tab:zNFW}, an increase in the cluster mass affects dramatically
the number of arcs, however, this increase just produces a minor shift of $N_{\rm arcs}$ towards lower redshifts.
This can be observed as a decrease of $z_{\rm cut}$ from 0.13 to 0.08 and $z_{\rm peak}$ from 0.68 to 0.55 as $\Mvir$ grows from 0.5 to 2 $M_{15}$ (see Table~\ref{tab:zNFW}).
From the upper panels in Figure~\ref{fig:NFWa}, and contrary to our expectations, we find that the statistics of arcs is almost independent on the definition of mass that we are using.\footnote{Under the assumption of a NFW profile, for a same halo $M_{200}$ is expected to be lower than \Mvir, since $r_{200} < r_{\Delta}$. Therefore, if we compare two haloes with the same value of mass, the first defined with $M_{200}$ and the second defined with \Mvir, we expect that $N_{\rm arcs}$ is higher in the first.} This result should be related to departures from the NFW shape in the $\Lambda$CDM simulations of \cite{2008MNRAS.390L..64D}. 
The curves in the lower left panel show that the NFW
distribution of arcs is strongly sensitive to the concentration parameter.
A higher \cvir\ will increase the number of arcs and shift
the distributions to lower redshifts. As shown in
Table~\ref{tab:zNFW}, changing the concentration parameter at
$z=0$ from $\cvir_0=8$ to $4$ increases the redshift where the
curves start to rise from $z_{\rm cut}=0.05$ to $0.16$, and where
the curves peak from $z_{\rm peak}=0.41$ to 0.78. 
From the lower right panel we find that an increase in the ellipticity produces a shift of the arc statistics toward lower redshifts. 
This is reflected in a slight decrease of $z_{\rm cut}$ from 0.10 to 0.06 and a strong decrease of $z_{\rm peak}$ from 0.60 to 0.42 as $\epsilon$ increases from 0.0 to 0.3 (see Table~\ref{tab:zNFW}).

For the NSIS case we find some important differences in the
$N_{\rm arcs}(z_L)$ curves when $r_c$ and $\epsilon$ are varied. In Figure~\ref{fig:NSISa} we plot $N_{\rm arcs}(z_L)$ for dark matter haloes with virial mass $\Mvir=10^{15}h^{-1}M_\odot$. The left panel of Figure~\ref{fig:NSISa} corresponds to three values of core radius at $z=0$; $r_{{\rm c}0}=0$, 12 and 24$\,h^{-1}$kpc ($\epsilon=0$). The right panel of Figure~\ref{fig:NSISa} corresponds $r_{{\rm c}0}=12\,h^{-1}$kpc  and three different values of ellipticity $\epsilon=0.00$, 0.15 and 0.30. 
In general $z_{\rm cut}$ and $z_{\rm peak}$ will be strongly
dependent on the core radius (\rc). This effect is clearly seen
in Figure~\ref{fig:NSISa} and Table~\ref{tab:zNSIS}. Increasing $\rc_0$
from 0 to $24\,h^{-1}$kpc will increase $z_{\rm cut}$ from 0.00 to 0.20
and $z_{\rm peak}$ from 0.00 to 0.57.  
Additionally, we find that increasing the ellipticity does not change $z_{\rm cut}$, but produces a preferential enhancement on the arc statistics at redshifts close to $z_{\rm cut}$. 
In particular, as seen in Figure~\ref{fig:NSISa} and Table~\ref{tab:zNSIS}, for a dark matter profile with $\Mvir=10^{15}h^{-1}M_\odot$, $r_{{\rm c}0}=12\,h^{-1}$kpc, when $\epsilon$ increases from 0.0 to 0.3,  $z_{\rm peak}$ decreases from 0.33 to 0.17.
\begin{figure}
   \includegraphics[width=8.4cm]{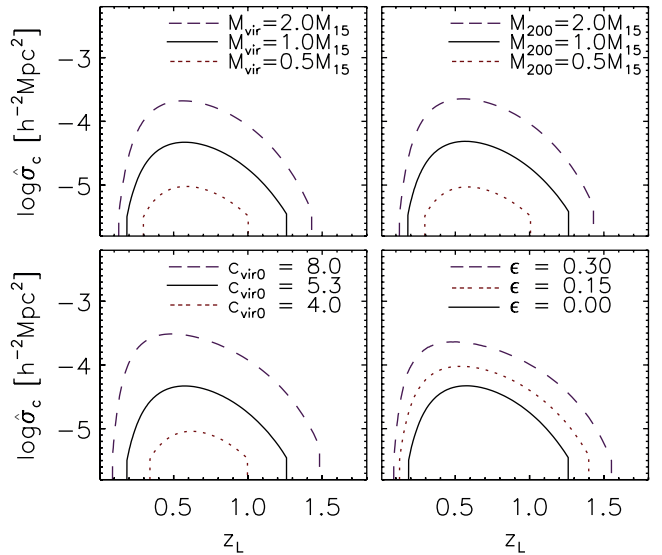}
        \centering
       \caption{Logarithm of NFW cross sections (in units of $h^{-2}{\rm Mpc^2}$) in function of lens redshift ($z_L$); the source redshift is fixed to $z_s=2.0$.
       Refer to legend in Figure~\ref{fig:NFWa} for details in the selection of parameters to estimate the curves in each panel.}
     \label{fig:NFWc0}
     \end{figure}
\begin{figure}
   \includegraphics[width=8.4cm]{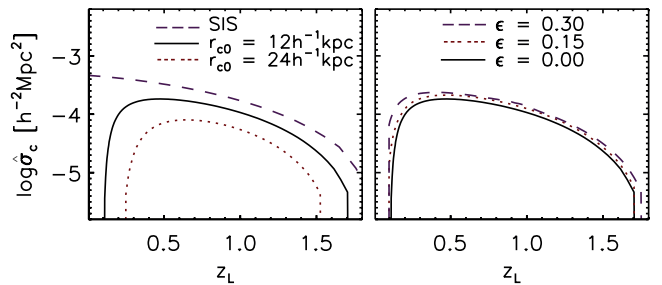}
        \centering
       \caption{Logarithm of NSIS cross sections (in units of $h^{-2}{\rm Mpc^2}$) in function of lens redshift ($z_L$); the source redshift is fixed to $z_s=2.0$.  Refer to legend in Figure~\ref{fig:NSISa} for details in the selection of parameters to estimate the curves in each panel.  }
     \label{fig:NSISc0}
     \end{figure}

\subsection{Strong lensing cross-section}
\label{cro}

From equation~\ref{eq:naog}, we expect that parameters derived from $N_{\rm arcs}(z_L)$ (like e.g., $z_{\rm cut}$ and $z_{\rm max}$) are mostly influenced by the lensing cross section.  
To show this effect we have calculated cross sections in the source plane with a fixed source redshift.
We have plotted the cross sections as a function of lens
redshift ($z_L$), in the NFW model (Figure~\ref{fig:NFWc0}) and NSIS
model (Figure~\ref{fig:NSISc0}) respectively. In both figures we have
assumed that the redshift of the source is $z_s=2$. 

As in the case
of the number of arcs (\S\ref{S:nar}), for NFW models, we have varied some
parameters of the lens halo (Figure~\ref{fig:NFWc0}) like
$M_\Delta$ (upper left panel), $M_{200}$ (upper
right panel), concentration parameter ($c_\Delta$) (lower right panel), and $\epsilon$ (lower left panel).
From the upper panels of Figure~\ref{fig:NFWc0}, we confirm that the effect of changing the mass in the NFW model is analogous to what we see in $N_{\rm arcs}$ (Figure~\ref{fig:NFWa}, \S\ref{S:nar}).
 In the lower left panel of Figure~\ref{fig:NFWc0}, we find that for a NFW halo with $M_{\rm vir}=10^{15}h^{-1}M_\odot$, the minimum redshift at which the cross sections become non-negligible grows 
from $0.09$ to $\sim 0.33$ and the redshift at which the cross sections are peaking grows from
$0.49$ to $0.62$ as $c_\Delta$ decreases from $8$ to $4$. 
From the lower right panel of Figure~\ref{fig:NFWc0}, we find that increasing the ellipticity decreases the minimum redshift at which the cross sections become non-negligible and the redshift at which the cross sections are peaking. 
Consequently, increasing ellipticity produces a preferential enhancement of the cross section close to redshifts where it starts to be non-negligible.
These
results are analogous to those found in the discussion of $N_{\rm
arcs}$ in function of $z_L$ depicted in Figure~\ref{fig:NFWa}.

 In the
case of the NSIS profiles, the cross sections (Figure~\ref{fig:NSISc0}) are strongly dependent on
the core radius ($r_c$). As $r_c$ increases, the cross sections
decrease in size and in redshift range. For a NSIS halo with $M_{\rm vir}=10^{15}h^{-1}M_\odot$, if $r_c$ increases
from 0 to $24\,h^{-1} {\rm kpc}$, the minimum redshift at which the
cross sections start to be non-negligible grows from $\sim 0.00$ to
$\sim 0.25$. Additionally, increasing the ellipticity will produce a preferential enhancement of the cross section at the redshift where they start to be non-negligible.
These results are similar to those in $N_{\rm arcs}(z_L)$ for NSIS models
(see Figure~\ref{fig:NSISa}  and \S\ref{S:nar}). For a fixed $z_s$ different
than two, the cross sections of both the NFW and NSIS profile will
have similar overall shape to those presented in
Figures~\ref{fig:NFWc0} and \ref{fig:NSISc0}. However, 
the maximum redshift at which these curves are non-negligible
will increase or decrease accordingly with a $z_s$ greater or
lower than two. Our cross sections
estimations (for $z_s=1$) are similar to those found in Figure~1
of \cite{2003MNRAS.340..105M}. These calculations (with $\Delta=200$)
  were confirmed by assuming a constant concentration
parameter $c_{200}$ in the NFW model and
$f_\sigma\sim0.7$ in the SIS model.\footnote{In
\cite{2003MNRAS.340..105M} the authors calculate cross sections using the SIS
(NSIS with $r_c=0$) and NFW model.}
\begin{figure}
   \includegraphics[width=8.4cm]{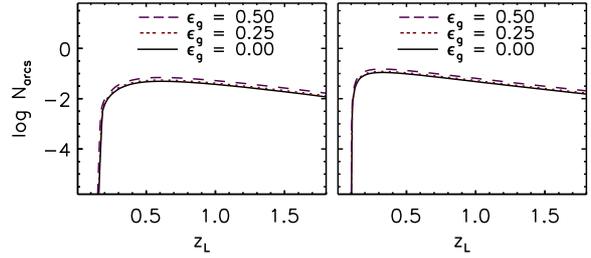}
        \centering
       \caption{Effect of the lensed galaxy ellipticity in in $N_{\rm arcs}$.
       The left/right panel correspond to the NFW and NSIS axially symmetric dark matter profiles. The curves in both panels are obtained with $M_{\rm vir}=10^{15} h^{-1} M_\odot$, $m_{\rm lim}=24$, $R_t=10$,  and three different lensed galaxies ellipticities $\epsilon_g=0.00$ (solid line), 0.25 (dotted line) and 0.50 (dashed line). In the left panel  \cvir\ is from \protect\cite{2008MNRAS.390L..64D} and in the right panel $\rco=12\,h^{-1}\,{\rm kpc}$ (the dependency $r_c$ with $z_L$ is given by equation~\ref{eq:rcmz}).}
     \label{fig:egal}
     \end{figure}

\subsection{Additional effects in the arc distribution.}\label{S:aarc}

In \S \ref{S:meto} we have assumed that the galaxy clusters (lenses) are a dark matter potential with elliptical symmetry. Additionally, as described in the beginning of this section, our calculations have been performed considering that the galaxies behind the lenses have finite size.
These simplifications minimize the number of free parameters required in our formulation. 
However, in order to find the limitations of this approach, we need to estimate the effect produced by some parameters that have been ignored. In this section, we estimate the variation of the number of arcs as a function of the ellipticity of the lensed galaxies based on  \cite{2001ApJ...562..160K}.  
In the end of this section, we also include comments about other parameters that could be relevant for our calculations. 

The dependency of the number of arcs on the ellipticity of the lensed galaxies is shown in Figure \ref{fig:egal}. In this Figure we present the number of arcs in function of $z_L$ for NFW (left panel) and NSIS (right panel) dark matter profiles with virial mass $\Mvir=M_{15}$. In each panel of this plot we have used three different values for the ellipticity of the lensed galaxies, $e_g=0.00$, 0.25 and 0.50.
From this figure we see a slight increase in the number of arcs as a function of the lensed galaxies ellipticity. 
From Figure~\ref{fig:egal} and Tables~\ref{tab:zNFW} and \ref{tab:zNSIS}, we conclude that increasing the lensed galaxy ellipticity increases the number of arcs, however it does not change the overall shape of $N_{\rm arcs}(z_L)$.

The effect of the seeing  will be to  circularize the object image, therefore larger seeing will decrease the number of arcs. This effect has been explored by \cite{2001AJ....121...10C}, and their conclusion was that for a seeing of $\gtrsim 1\farcs5$ we expect a decrease in the $N_{\rm arcs}$ by a factor close to one order of magnitude. 
In a survey where the seeing is $\lesssim 1\farcs0$ and is not varying significantly between observations, we expect that this effect will not be important in affecting the shape and amplitude of the distribution of the number of arcs. 

There are other effects that could be affecting the statistics of arcs that are out of the scope of this paper. Among them are: the triaxiality on the lens mass distribution  \citep{2003ApJ...599....7O}, cluster mergers \citep{2004MNRAS.349..476T}, halo concentration distribution \citep{2007A&A...473..715F},   cluster asymmetries and substructures \citep{2007A&A...461...25M}, influence of stellar mass in galaxies \citep{2008MNRAS.386.1845H},  and baryonic cooling \citep{2008ApJ...676..753W, 2008ApJ...687...22R}. 
Although  most of these effects will affect substantially the lens statistics, it is expected that the shape of $N_{\rm arcs}(z_L)$ will be robust for low redshift clusters  ($z_L \lesssim 0.3$). This is because at low redshift $N_{\rm arcs}(z_L)$ is mostly dependent on the encircled mass near the center of the mass distribution of the lens.  
Notice though that the baryonic matter contribution could be important in the inner regions of the clusters. However, latest studies suggest that pure dark matter models (like the ones used in this work) could be enough to describe cluster haloes even in those regions where baryonic matter is important \citep{2013ApJ...765...25N,2013ApJ...765...24N}.
Future work with more complex lens models will be helpful to understand better the power of the simplified methodology used in this work.  In the next section, we briefly describe how to apply this formulation to an ensemble of low redshift galaxy clusters. Through this approach, we expect to obtain first order constrains on parameters that have an impact on the encircled mass at $r\ll r_\Delta$. 
  \begin{figure}
   \includegraphics[width=8.4cm]{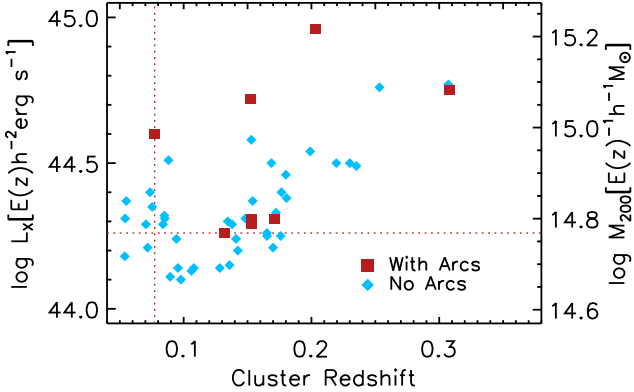}
        \centering
       \caption{$L_{\rm X}$ as a function of redshift for our bright sample of Abell clusters. Squares and diamonds are cluster with and without evidence of arcs respectively. Vertical and horizontal dotted lines are used to mark the minimum redshift and minimum X-ray luminosity of clusters that present gravitational arcs.}
     \label{Abell}
     \end{figure}

\subsection{Kolmogorov-Smirnov test  to an ensemble of galaxy clusters} \label{S:appl}

To give applicability to our method, the mass of the cluster ($M_\Delta$) must be related to some observable parameter.  For this purpose, we use the empirical relation  between the X-ray luminosity of the clusters and $M_{200}$ (i.e., $\Delta=200$) obtained by \cite{2010ApJ...709...97L}, which can be expressed as 
\begin{equation}  \label{eq:M200}
M_{200}(L_{\rm X},z)\!=\!\frac{B^*}{E(z)}\left(\frac{L_{\rm X}E(z)^{-1}}{10^{44}\,h^{-2}\,\lumin}\right)^{A^*} 10^{15}h^{-1}M_\odot,
\end{equation}
where $L_{\rm X}$ is the X-ray luminosity in the 0.1--2.4 keV band, $A^*\approx 0.64$, and $B^*\approx 0.400$.\footnote{The original expression presented in \cite{2010ApJ...709...97L} is $M_{200}E(z)=M_0 {B}(L_{\rm X}E(z)^{-1}/L_{X0})^{A}$, where $M_0=10^{13.7}h_{72}^{-1}\,M_\odot$, $L_{X0}=10^{42.7}h_{72}^{-2}\, \lumin$, and  $h_{72}$ is $H_0$ in units of 72~$\kms {\rm Mpc}^{-1}$. The values of $A^*\approx0.64$, and $B^*\approx0.400$ in equation~\ref{eq:M200} are obtained rewriting this equation with $A=0.64$ and ${\rm log}\,B=0.03$.} 
\begin{table*}
\begin{minipage}{175mm}
\begin{center}
\caption{Abell Clusters observed with VLT  \label{tab:abel}}
\begin{tabular}{ccrccccccccccccc}
\hline\hline
{\sc object name} &
$\alpha_{2000.0}$ &
$\delta_{2000.0}$ &
$z$ &
{\sc ref}  &
$f_{\rm X}$ &
${\rm log}\,L_{\rm X}$ &
${\rm log}\,M_{200}$ &
Arcs? \\
(1) &
(2) &
(3)~~ &
(4) &
(5)  &
(6) &
(7) &
(8) &
(9) \\
\hline
     A0022  &    5.161  &  $-25.7220$  &    0.1424  &  8  &   7.3  &     44.20  &     14.73  &  no  \\
     A0085  &   10.453  &  $-9.31800$  &    0.0551  &  5  &  72.3  &     44.37  &     14.84  &  no  \\
     A0141  &   16.388  &  $-24.6500$  &    0.2300  &  2  &   5.7  &     44.50  &     14.92  &  no  \\
     A0399  &   44.457  &   $13.0530$  &    0.0718  &  5  &  29.0  &     44.21  &     14.73  &  no  \\
     A0401  &   44.737  &   $13.5730$  &    0.0737  &  5  &  42.6  &     44.40  &     14.86  &  no  \\
     A0478  &   63.359  &   $10.4660$  &    0.0881  &  2  &  39.1  &     44.51  &     14.93  &  no  \\
     A0520  &   73.531  &   $2.92000$  &    0.1990  &  2  &   8.3  &     44.54  &     14.95  &  no  \\
     A0545  &   83.097  &  $-11.5360$  &    0.1540  &  2  &   9.2  &     44.37  &     14.84  &  no  \\
     A0644  &  124.355  &  $-7.52800$  &    0.0704  &  2  &  36.8  &     44.29  &     14.79  &  no  \\
     A0750  &  137.299  &   $10.9890$  &    0.1800  &  2  &   8.4  &     44.46  &     14.90  &  no  \\
     A0754  &  137.256  &  $-9.65500$  &    0.0542  &  2  &  64.1  &     44.31  &     14.80  &  no  \\
     A0780  &  139.528  &  $-12.0990$  &    0.0539  &  2  &  48.4  &     44.18  &     14.72  &  no  \\
     A0795  &  141.024  &   $14.1680$  &    0.1359  &  2  &   7.1  &     44.15  &     14.70  &  no  \\
     A0901  &  149.122  &  $-9.94800$  &    0.1700  &  4  &   5.2  &     44.21  &     14.73  &  no  \\
     A0907  &  149.589  &  $-11.0610$  &    0.1527  &  1  &   8.1  &     44.31  &     14.80  & yes  \\
     A1084  &  161.128  &  $-7.08400$  &    0.1323  &  8  &   9.7  &     44.26  &     14.77  & yes  \\
     A1285  &  172.586  &  $-14.5750$  &    0.1061  &  2  &  11.2  &     44.13  &     14.69  &  no  \\
     A1300  &  172.979  &  $-19.9140$  &    0.3072  &  2  &   6.1  &     44.77  &     15.10  &  no  \\
     A1437  &  180.106  &   $3.35100$  &    0.1345  &  8  &  10.2  &     44.30  &     14.79  &  no  \\
     A1451  &  180.811  &  $-21.5270$  &    0.1711  &  1  &   6.5  &     44.31  &     14.80  & yes  \\
     A1553  &  187.700  &   $10.5560$  &    0.1652  &  2  &   6.1  &     44.25  &     14.76  &  no  \\
     A1650  &  194.674  &  $-1.75600$  &    0.0838  &  8  &  25.6  &     44.29  &     14.79  &  no  \\
     A1651  &  194.850  &  $-4.18900$  &    0.0849  &  8  &  27.1  &     44.32  &     14.81  &  no  \\
     A1664  &  195.934  &  $-24.2560$  &    0.1283  &  8  &   7.8  &     44.14  &     14.69  &  no  \\
     A2029  &  227.729  &   $5.72000$  &    0.0773  &  5  &  61.6  &     44.60  &     14.98  & yes  \\
     A2104  &  235.027  &  $-3.30600$  &    0.1533  &  8  &   7.7  &     44.29  &     14.79  & yes  \\
     A2163  &  243.956  &  $-6.15000$  &    0.2030  &  2  &  21.0  &     44.96  &     15.22  & yes  \\
     A2204  &  248.195  &   $5.57400$  &    0.1522  &  8  &  21.2  &     44.72  &     15.06  & yes  \\
     A2345  &  321.744  &  $-12.1410$  &    0.1765  &  2  &   7.6  &     44.40  &     14.86  &  no  \\
     A2384  &  328.069  &  $-19.6000$  &    0.0943  &  2  &  18.2  &     44.24  &     14.76  &  no  \\
     A2426  &  333.635  &  $-10.3650$  &    0.0978  &  2  &  12.2  &     44.10  &     14.67  &  no  \\
     A2597  &  351.319  &  $-12.1240$  &    0.0852  &  2  &  25.9  &     44.31  &     14.80  &  no  \\
     A2744  &    3.567  &  $-30.3830$  &    0.3080  &  2  &   5.7  &     44.75  &     15.08  & yes  \\
     A2811  &   10.533  &  $-28.5360$  &    0.1079  &  9  &  10.9  &     44.14  &     14.69  &  no  \\
     A3017  &   36.485  &  $-41.9060$  &    0.2195  &  3  &   6.2  &     44.50  &     14.92  &  no  \\
     A3041  &   40.333  &  $-28.6870$  &    0.2352  & 10  &   5.3  &     44.49  &     14.92  &  no  \\
     A3112  &   49.485  &  $-44.2380$  &    0.0753  &  9  &  36.4  &     44.35  &     14.82  &  no  \\
     A3292  &   72.459  &  $-44.6860$  &    0.1723  &  1  &   6.8  &     44.33  &     14.82  &  no  \\
     A3364  &   86.906  &  $-31.8720$  &    0.1483  &  3  &   8.6  &     44.31  &     14.80  &  no  \\
     A3378  &   91.470  &  $-35.3010$  &    0.1410  &  1  &   8.2  &     44.24  &     14.76  &  no  \\
     A3396  &   97.205  &  $-41.7250$  &    0.1759  &  3  &   5.4  &     44.25  &     14.76  &  no  \\
     A3411  &  130.475  &  $-17.4930$  &    0.1687  &  6  &  10.5  &     44.50  &     14.92  &  no  \\
     A3444  &  155.953  &  $-27.2640$  &    0.2533  &  2  &   8.6  &     44.76  &     15.09  &  no  \\
     A3695  &  308.694  &  $-35.8300$  &    0.0894  &  2  &  15.1  &     44.11  &     14.67  &  no  \\
     A3739  &  316.073  &  $-41.3390$  &    0.1651  &  7  &   6.2  &     44.26  &     14.77  &  no  \\
     A3856  &  334.656  &  $-38.8870$  &    0.1379  &  2  &   9.5  &     44.29  &     14.79  &  no  \\
     A3888  &  338.637  &  $-37.7330$  &    0.1529  &  8  &  15.2  &     44.58  &     14.97  &  no  \\
     A3984  &  348.907  &  $-37.7480$  &    0.1805  &  2  &   6.9  &     44.38  &     14.84  &  no  \\
     A4010  &  352.809  &  $-36.5020$  &    0.0955  &  9  &  14.1  &     44.14  &     14.69  &  no  \\

\hline                                                                                            
\end{tabular}
\end{center}

Col. (1): cluster name. Cols (2) and (3): optical positions in J2000.0 equatorial coordinates. 
Col. (4): redshift. Col. (5): redshift reference: 1$\equiv$\cite{1996MNRAS.281..799E}; 
2$\equiv$\cite{1999ApJS..125...35S}; 3$\equiv$\cite{1999ApJ...514..148D};  
4$\equiv$\cite{2000A&AS..142..433S}; 5$\equiv$\cite{2001AJ....122.2858O};  
6$\equiv$\cite{2002ApJ...580..774E}; 7$\equiv$\cite{2004A&A...425..367B};  
8$\equiv$\cite{2006MNRAS.366..645P}; 9$\equiv$\cite{2006ApJ...638..725Z}; 
10$\equiv$\cite{2009AJ....137.4795C}  
Col. (6): 0.1--2.4~keV absorption corrected flux in units of $10^{-12}\,\flux$ 
\citep[from][]{1996MNRAS.281..799E}. 
Col. (7): ${\rm log}\,L_{\rm X}$, where $L_{\rm X}$  is the 0.1--2.4~keV luminosity 
in units of $E(z)\,h^{-2}\lumin$. 
Col. (8): ${\rm log}\,M_{200}$, where $M_{200}$ is in units of $E(z)^{-1}h^{-1}M_\odot$. 
Col. (9): yes if the cluster presents evidence of strong lensing images, no otherwise 
(see Appendix~\ref{S:lenc} for details).
\end{minipage}
\end{table*}


We apply our results to a real case by compiling the information of masses and presence of arcs in a large sample of galaxy clusters. 
We selected bright X-ray Abell clusters ($\lx > 1.2 \cdot 10^{44} h^{-2} \lumin$ or $M_{200} \gtrsim 5\cdot 10^{14}h^{-1}M_\odot$) in the Southern Hemisphere ($-50^\circ \leqslant \delta \leqslant 15^\circ$) with $z\geq0.05$.
The clusters were observed with the FORS1 instrument mounted on the VLT-Antu telescope. The requirement that $z\geq0.05$ was to ensure that a large fraction of the clusters fits inside the FOV of the camera ($6\farcm8\times6\farcm8$).
The observations were obtained under homogeneous sky conditions and sub-arcsecond image quality (median of 0\farcs6). 
The complete sample consists of 49 clusters (see Table~\ref{tab:abel}) and the weak lensing properties of 24 of them have been previously presented in \cite{2004ApJ...613...95C} (see here also for details on the data reduction).
The pixel scale used was $0\farcs2$ and the FOV length of 6\farcm8 corresponds to proper distances of $\approx 0.3$, 0.7 and 1.3~$h^{-1}\,{\rm Mpc}$ for small ($z=0.05$), average ($z=0.14$), and large ($z=0.3$) clusters redshifts. 
The {\emph V}, {\emph R}, {\emph I} bands imaging was centered on the cluster cores and with exposures times of 330~s in each filter. 
In our search of strong lensing images we found that 8 out of 49 clusters show strong lensing arcs (see Table~\ref{tab:abel} and Appendix~\ref{S:lenc}) and Figure~\ref{Abell}; the minimum redshift at which we found arcs was $\approx 0.08$.
 \begin{figure}
   \includegraphics[width=8.4cm]{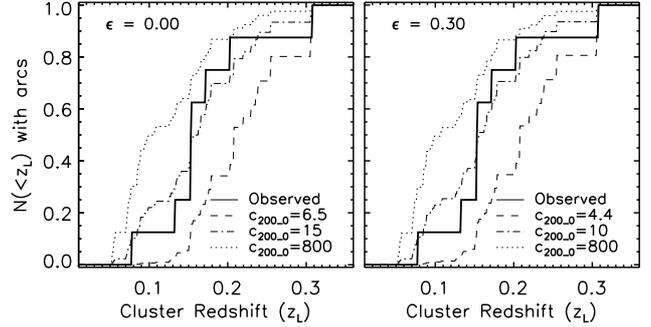}
        \centering
       \caption{Observed cumulative distribution of a sample of Abell clusters with arcs as a function of cluster redshift (solid line in both panels).  Left panel: axially symmetric NFW models ($\epsilon=0$) with concentration parameters of $\ctho=6.5$ (dashed line), 15 (dash-dotted line) and 800 (dotted line).  Right panel: NFW pseudo-elliptical models with $\epsilon=0.3$ and concentration parameters of $\ctho=4.4$ (dashed line), 10 (dash-dotted line) and 800 (dotted line). In all insets $m_{\rm lim}=24$, $R_t=10$, and the dependency of $c_{200}$ on $M_{200}$ and $z_L$ is given by equation~\ref{eq:ctho}.}
     \label{CNFW}
     \end{figure}
\begin{figure}
   \includegraphics[width=8.4cm]{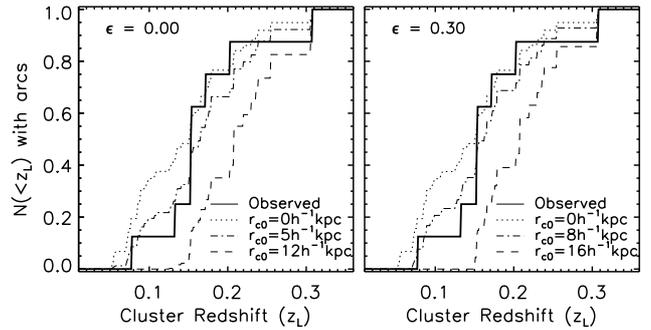}
        \centering
       \caption{Observed cumulative distribution of a sample of Abell clusters with arcs as a function of cluster redshift (solid line both panels). Left panel: axially symmetric NSIS models ($\epsilon=0$) with core radii of $\rco\,[{\rm h^{-1} kpc}]=0$ (SIS; dotted line), 5 (dash-dotted line) and 12 (dashed line).  Right panel: NSIS pseudo-elliptical models with $\epsilon=0.3$ and core radii of $\rco\,[{\rm h^{-1} kpc}]=0$ (SIS: dotted line), 8 (dash-dotted line) and 16 (dashed line).  In all insets $m_{\rm lim}=24$, $R_t=10$, and the dependency of $\rc$ on $M_{200}$ and $z_L$ is given by equation~\ref{eq:rcmz}.}
     \label{CNSIS}
     \end{figure}

We estimate the observed cumulative number of clusters with arcs on our sample as a function of their redshift, and compare this with the expected cumulative number of clusters with arcs from our models. 
 Based on \cite{2008MNRAS.390L..64D} for virialized clusters, in the case of the NFW model, we estimate the clusters concentration parameters using equation~\ref{eq:ctho} with $\alpha_M=-0.091$ and $\alpha_z=-0.44$.  In the case of the NSIS model, we parametrize the core radii using equation~\ref{eq:rcmz} with $M_{\rm vir}=(200/\Delta_{\rm vir}(z_L))^{1/2}M_{200}$.
 The assumed \Iband\ limiting magnitude is $m_{\rm lim}=24$, which is close to the sensitivity limit of our observations. However, the estimated cumulative number of clusters with arcs is insensitive to the limiting magnitude in the range $22 \lesssim m_{\rm lim} \lesssim 26$. The assumed value of $R_t$ is 10 albeit the cumulative distributions  are also not affected for $5 \lesssim R_t \lesssim 20$. Similar conclusions are obtained when we vary the ellipticity of the lensed galaxies from $\epsilon_g=0$ to $\epsilon_g=0.5$. In general, the cumulative distributions should be not affected by any additional parameter that does not change appreciably the shape of $N_{\rm arcs}(z_L)$.

 The observed cumulative number of clusters with arcs of our Abell cluster sample has been compared with six different prescriptions of each NFW model (Figure~\ref{CNFW}) and NSIS model (Figure~ \ref{CNSIS}) respectively.
 The three model curves presented in the left panel of Figure~\ref{CNFW} correspond to axially symmetric NFW profiles with $c_{200\_0}=$~6.5, 15 and 800 respectively. The three model curves shown in the  right panel Figure~\ref{CNFW} correspond to pseudo-elliptical NFW profiles with $\epsilon=0.3$ and $c_{200\_0}=$~4.4, 10 and 800 respectively.
Notice that a value of the pseudo-elliptical ellipticity $\epsilon=0.3$ correspond to a value of $\epsilon_\Sigma\sim 0.5$ (see section \S\ref{Pseudo} and Appendix~\ref{S:apps}). This particular value of $\epsilon_\Sigma$ is close to estimations based on weak and strong lensing of the sample presented in \cite{2012MNRAS.420.3213O}. 
The three model curves shown in the left panel of Figure~\ref{CNSIS} correspond to axially symmetric NSIS profiles with $r_{\rm c0}=$~0 (SIS), 5 and 12 $h^{-1}$kpc respectively. The three model curves presented in the right panel Figure~\ref{CNSIS} correspond to pseudo-elliptical NSIS profiles with $\epsilon=0.3$ with $r_{\rm c0}=$~0 (SIS), 8 and 16~$h^{-1}$kpc respectively.
  
  The KS test is used to verify if the null hypothesis that the observed number of arcs comes from the same distribution than those estimated from the models. 
In the panels of Figure~\ref{CNFW}, the curves with extreme values of $\ctho$ correspond to cases where the predicted cumulative distributions depart significantly from the  observed ($\approx 95\%$ level of confidence). In the panels of  Figure~\ref{CNSIS} the only curves that depart significantly form the observed are those with the maximum value of $r_{\rm c0}$. 
In consequence, we conclude that we  reject the null hypothesis at a confidence level \hbox{$>95\%$} for $\ctho \lesssim 6.5$ and $\ctho \gtrsim 800$ in cases of NFW profiles without ellipticity and for $\ctho \lesssim 4.4$ and $\ctho \gtrsim 800$ in cases of NFW profiles with ellipticity ($\epsilon=0.3$). 
Additionally, we  reject the null hypothesis at a confidence level $>95\%$ for $r_{\rm c0} \gtrsim 12 h^{-1}$~kpc and for $r_{\rm c0} \gtrsim 16 h^{-1}$~kpc in cases of NSIS profiles without and with ellipticity ($\epsilon=0.3$) respectively. 
Notice that although incrementing the ellipticity does not significantly affect $z_{\rm cut}$,  
it preferentially enhances the statistics of clusters at low redshifts (as described in \S\ref{S:nar}). 
Therefore, when we compare axially symmetric models to
 cases where ellipticity is taken into account,
we obtain clear differences on the constraints on $\ctho$ or $r_{\rm c0}$.
Based of expressions~\ref{eq:zcNFW}  and \ref{eq:zcNSIS} the estimations of $\ctho$ and $r_{c0}$ allow to calculate  approximate values of $z_{\rm cut}$ for clusters with masses within those on our sample.
If we assume $M_{200}=10^{\langle {\rm log}\,M_{200} \rangle}\approx0.62\,M_{15}$ with $\langle {\rm log}\,M_{200} \rangle$ the logarithm  averaged cluster mass,  we find that $z_{\rm cut} \lesssim 0.07 $ for NFW profiles and $z_{\rm cut} \lesssim 0.12$ for NSIS.

As a reference, the  results from \lcdm\ simulations of \cite{2008MNRAS.390L..64D} indicate that $\ctho \approx 3.8$ ($\sigma_{\rm log c}\sim 0.12$).
Hence, although we find more concentrated haloes ($\ctho \gtrsim 4.4$) than those of \cite{2008MNRAS.390L..64D}, our expected values of $\ctho$ fall within the predicted errors of the simulations.
The concentrations found in this work are also consistent with $c_{200\_0}\sim 5.4$ as predicted by $N$-body \lcdm\ simulations of  \cite{2012MNRAS.423.3018P}.\footnote{\cite{2012MNRAS.423.3018P} estimate higher concentration parameters at masses in the range of galaxy clusters than previous \lcdm\ simulations. This result is attributed to the fact that \cite{2012MNRAS.423.3018P} find little evolution on massive haloes, and as a consequence, higher concentrations than previous 
studies.}
Note that larger strong lensing surveys on  clusters at low redshift can provide better constraints on cumulative distribution of clusters with arcs, and thus, more restrictive values on $\ctho$, $r_{c0}$ and $z_{\rm cut}$.
For example, including clusters with $z\lesssim 0.05$ in our sample should decrease our estimations on the upper limit of $
\ctho$ for NFW profiles and/or increase the lower limit of $r_{\rm c0}$ for NSIS profiles.

As mentioned in \S \ref{S:aarc}, there are many effects not considered in our approach that could have an influence on our results.
In particular, selecting a sample of massive (X-ray bright) and low redshift clusters expected to be mostly relaxed and virialized, and therefore, with more concentrated mass profiles than normal populations of galaxy clusters \citep[e.g.,][]{2013ApJ...776...39R}.
Additionally, strong lensing should be preferably observed  in clusters where the line of sight is oriented along the main axis of their triaxial mass profiles \citep[e.g.,][]{2004MNRAS.350.1038C, 2005A&A...443..793G,2005ApJ...632..841O}.
The selection effects in consideration could be artificially increasing the observed concentration up to 30\% \citep[e.g.,][]{2012MNRAS.420.3213O, 2014ApJ...797...34M, 2015ApJ...806....4M}. 
Considerations of these type should be analyzed in future studies by using more complex models than those in this work.

\section{Summary and conclusions} \label{S:conc}

In this paper we introduce an axially symmetric formula
(equation~\ref{eq:naog}) to calculate the probability of finding
strong lensing  arcs in galaxy clusters as a function of
their redshift and virial mass. 
This formula has been modified in order to  include ellipticity through the use of a pseudo-elliptical approximation.  We have tested this formulation
using the NFW and NSIS dark matter mass profiles, and we have studied its
dependency on the mass, core
radius, concentration parameter, and ellipticity.  

For the NFW profiles, we have confirmed
that the halo cluster masses produce important variation on the
number of arcs detected. Incrementing by a factor of four the halo cluster mass will increase the number of arcs in approximately an order of magnitude.  However, such increment in mass does not change significantly the shape of the number of arcs distributions as a function of redshift.
In this model, changes in the concentration parameter
produce substantial variations in the number of arcs as a function
of the cluster redshift. In particular, for a
NFW profile with virial mass $\sim10^{15} h^{-2} M_\odot$, a
change in the concentration parameter normalization ($c_{\Delta0}$) from 4 to 8
will shift the minimum cluster redshift where we find arcs from
$z_{\rm cut}\sim0.16$ to 0.05. Such change will also  vary the redshift where the
$N_{\rm arcs}$ are maximum from $z_{
\rm peak}\sim 0.8$ to 0.4.  

In the case of the NSIS models, the distribution of the
arcs is very sensitive to the core radius
of the model. In particular, for a NSIS profile
with virial mass $\sim10^{15} h^{-2} M_\odot$, a core radius
ranging from 0 to 24~$h^{-1}$kpc will produce dramatic variations
on the lens statistics. For this case, the minimum redshift where
we find arcs is shifted from $z_{\rm cut}\sim0.0$ to 0.2, and the
redshift where the $N_{\rm arcs}$ are maximum vary from $z_{\rm
peak}\sim0.0$ to 0.6.
\newline In both dark matter profiles studied in this work, we find that
an increase in ellipticity does  not significantly change $z_{\rm cut}$, however, 
it produces a preferential enhancement on the arc statistics at redshifts close 
to $z_{\rm cut}$.
This effect is clearly observed through a strong decrease of $z_{\rm peak}$ as the ellipticity
of the dark matter potential increases. Therefore, the ellipticity has an important
impact on the statistics of arcs.

We have implemented our method to analyze the arc statistics of an \XR\ bright low redshift sample of Abell clusters that were observed with VLT. Through a simple KS test, we have been able to constrain the concentration parameter for NFW profiles and the core radius for NSIS profiles.
For  NFW profiles, we obtain that $\ctho \geq 6.5$ for axially symmetric profiles, and  $\ctho \geq 4.4$ for elliptical profiles. 
Consequently, our estimations of the concentration are consistent with those predicted by $N$-body \lcdm\ simulations (within the rms errors).
Additionally, for NSIS profiles, our Abell cluster arc statistics provide upper limits on the core radius ($r_{\rm c0} \leq 16\,h^{-1} {\rm kpc}$) and thus a SIS model is not ruled out from our observations. 

For an arbitrary density profile, our approach should be useful to estimate parameters that modify the mass distribution near the halo center ($r\ll r_\Delta$). 
These parameters are related with the lowest cluster redshift where strong arcs can be observed ($z_{\rm cut}$) for a well defined sample of galaxy clusters.  Such lowest redshift is expected
to lie in the range 0.0--0.2, highlighting the need to adopt a very low-$z$ limit for samples to study the clusters mass profiles.

\section{Acknowledgements}
We thank the anonymous referee for his/her insightful comments that helped us improve this work. 
JPK acknowledges support from the ERC advanced grant LIDA and from CNRS.
LSJ is partially supported by FAPESP (project 2012/00800-4) and CNPq.  
ESC is partially supported by FAPESP ( 2014/13723-3) and CNPq.
LEC received partial support from the Center of Excellence in Astrophysics and Associated Technologies (PFB06
) and from CONICYT Anillo project ACT-1122.
CS acknowledges support from CONICYT-Chile (FONDECYT 3120198 and Becas Chile 74140006). 
Based on observations made with ESO Telescopes at the La Silla Paranal Observatory under programme IDs 67.A-0597(A) and 70.B-0440(A).


\bibliographystyle{apj}

\providecommand{\noopsort}[1]{}

\appendix

\section{Comparison of arc statistics of radial images versus tangential images} \label{S:aprt}
\begin{figure}
   \includegraphics[width=8.4cm]{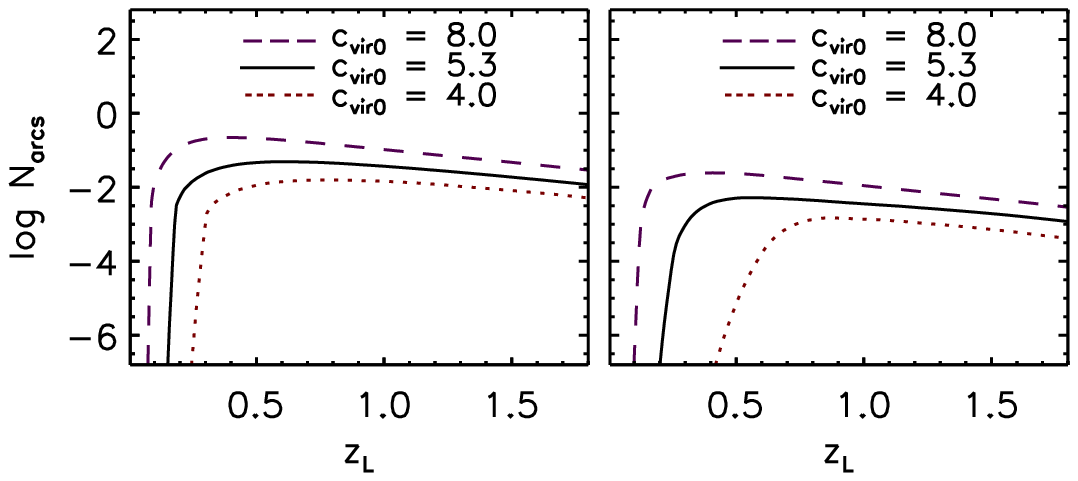}
        \centering
     \caption{Tangential (left panel) and radial (right panel) number of arcs as a function of $z_L$ in NFW profiles. The left panel corresponds to the lower left panel of Figure~\ref{fig:NFWa} but with different scaling. The parameters to obtain the radial number of arcs in the right panel are identical to those of the left panel. Refer to legend in Figure~\ref{fig:NFWa} for details in the selection of the lensing parameters.}
     \label{NFWrad}
     \end{figure}
\begin{figure}
   \includegraphics[width=8.4cm]{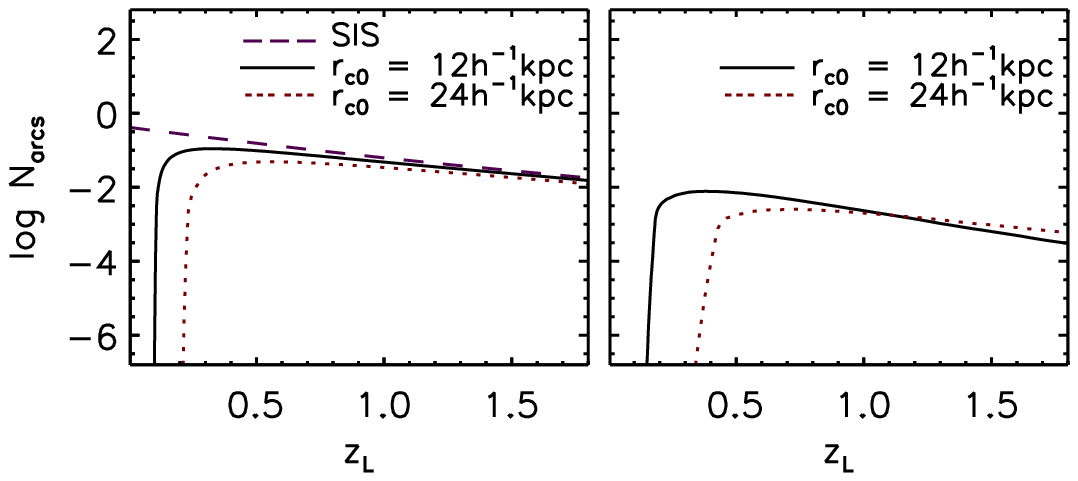}
        \centering
     \caption{Tangential (left panel) and radial (right panel) number of arcs as a function of $z_L$ in NSIS profiles. 
The left panel corresponds to the left panel of Figure~\ref{fig:NSISa} but with different scaling. The parameters to obtain the radial number of arcs in the right panel are identical to those of the left panel. The absence of a SIS curve in the right panel is due to the lack of radial images in this case.  Refer to legend in Figure~\ref{fig:NSISa} for details in the selection of the lensing parameters.}
     \label{NSISrad}
     \end{figure}
We proceed to obtain the number of arcs using equation~\ref{eq:naog} for radial images in axially symmetric profiles. 
In Figure~\ref{NFWrad} left panel we have reproduced (using different axis scaling) the lower left panel of Figure~\ref{fig:NFWa}.
In the right panel of Figure~\ref{NFWrad} we obtain the number of radial arcs for NFW profiles using the same combination of parameters that the left panel. By comparing the left and right panel of Figure~\ref{NFWrad}, we find that tangential arcs are about one order of magnitude more numerous than radial arcs. 
Note that similar results are obtained if we compare the number of tangential and radial arcs for axially symmetric NSIS profiles (see Figure~\ref{NSISrad}).

\section{Pseudo elliptical approximation limits of validity} \label{S:apps}

\begin{figure}
   \includegraphics[width=8.4cm]{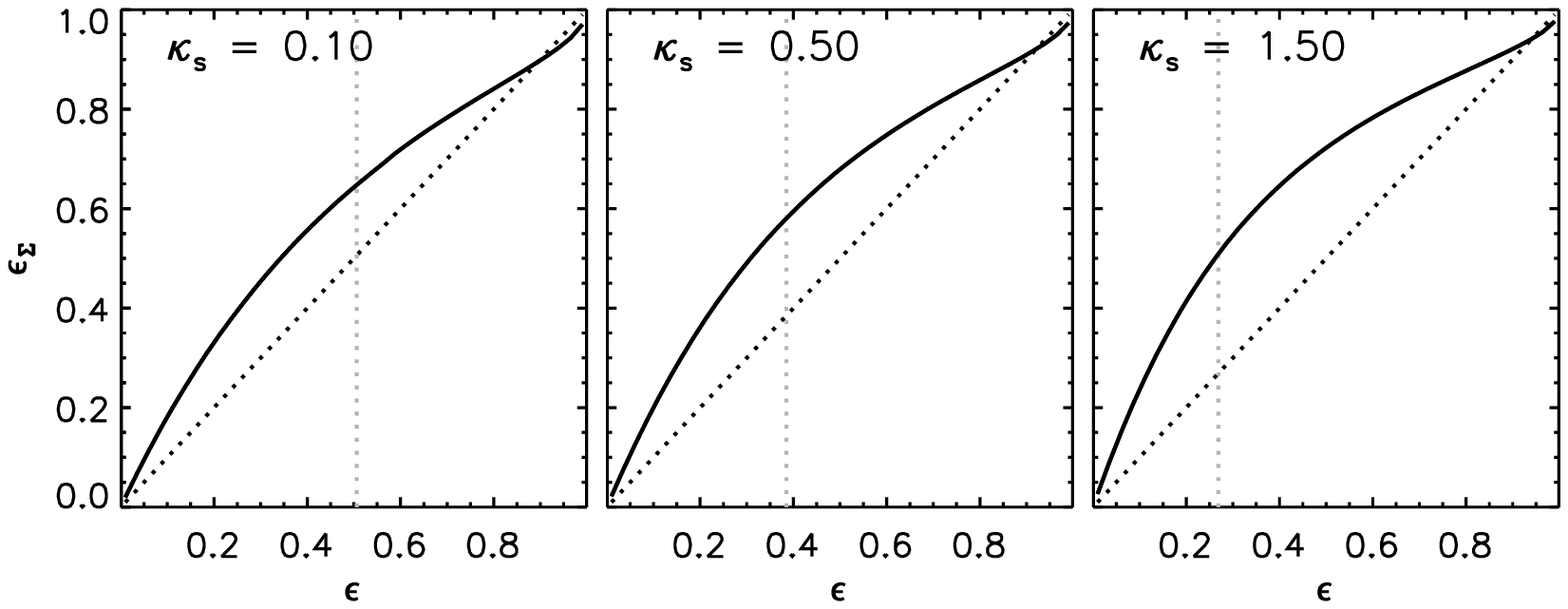}
        \centering
     \caption{Fitted ellipticity ($\epsilon_\Sigma$) of the pseudo elliptical mass distribution as a function of $\epsilon$. Each  panel corresponds to NFW models with three different values of $\kappa_s$: in left panel $\kappa_s=0.1$, in central panel $\kappa_s=0.5$, and in right panel $\kappa_s=1.5$. In each panel, the dotted diagonal line shows as reference $\epsilon_\Sigma = \epsilon$. Additionally, in each panel, the vertical dotted line indicates the maximum values of $\epsilon$ for which the pseudo elliptical function is acceptable ($\mathcal{D}^2 = 4.5\cdot10^{-4}$, where $\mathcal{D}^2$ is given in equation \ref{eq:Dmin}).}
     \label{epmxnfw}
     \end{figure}
\begin{figure}
   \includegraphics[width=8.4cm]{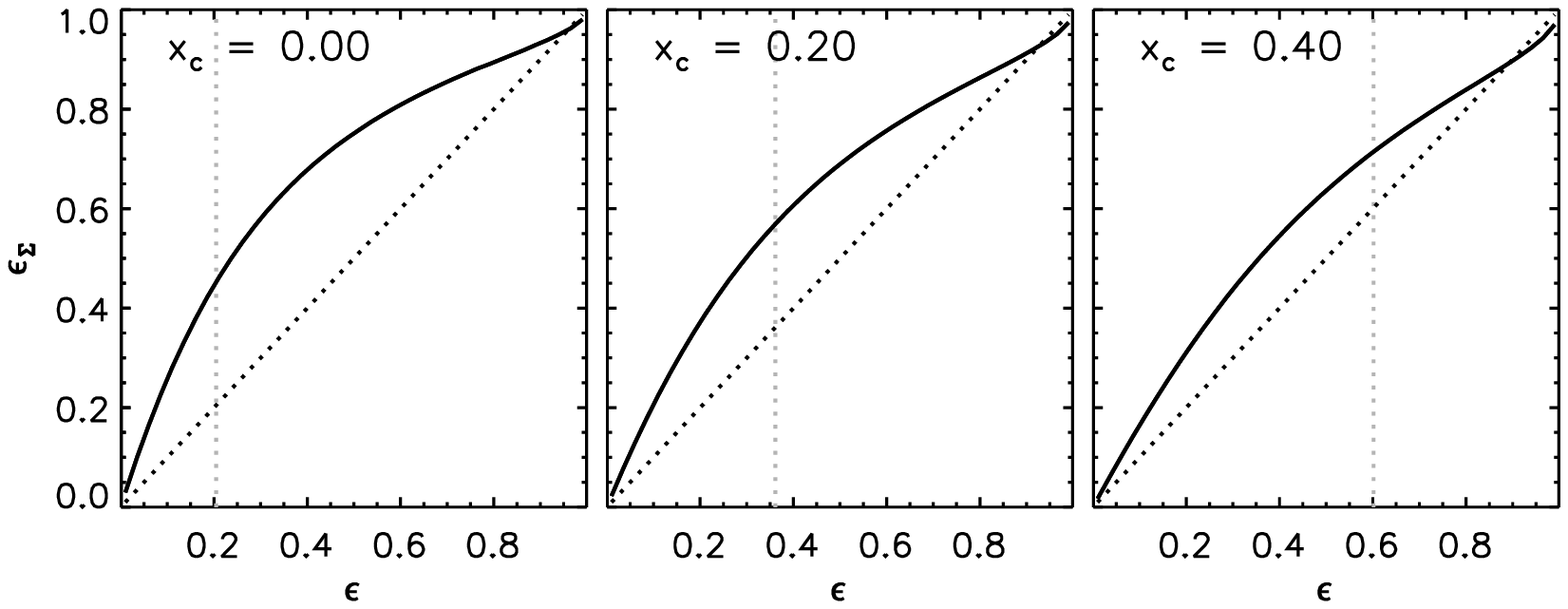}
        \centering
     \caption{Fitted ellipticity ($\epsilon_\Sigma$) of the pseudo elliptical mass distribution as a function of $\epsilon$. Each  panel corresponds to NSIS models with three different values of $x_c$: in left panel $x_c=0.0$ (SIS), in central panel $x_c=0.2$, and in right panel and $x_c=0.4$. For reference about dotted lines see legend of figure \ref{epmxnfw}.}
     \label{epmxnsis}
     \end{figure}
\begin{figure}
   \includegraphics[width=8.4cm]{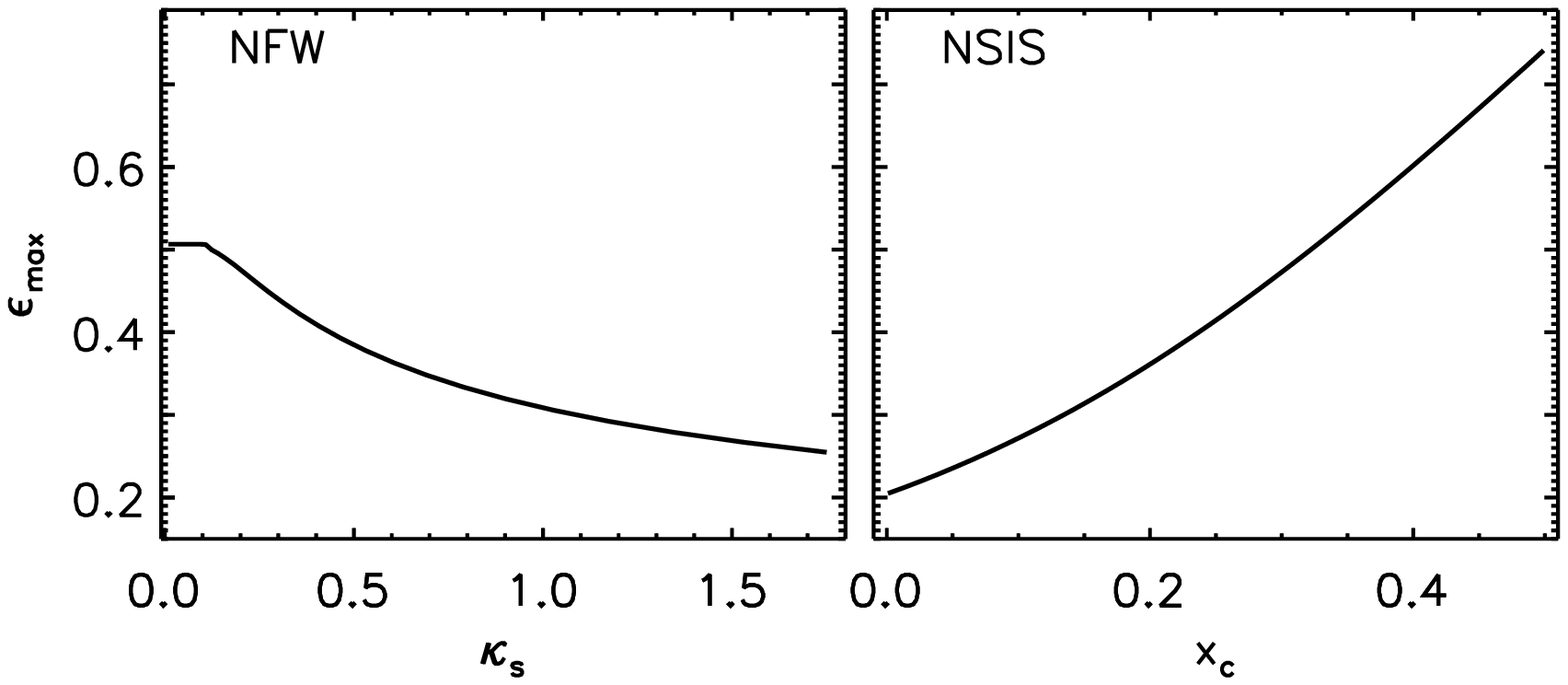}
        \centering
     \caption{Maximum value of the ellipticity for which the pseudo-elliptical approximation is acceptable ($\mathcal{D}^2 = 4.5\cdot10^{-4}$, where $\mathcal{D}^2$ is given in equation \ref{eq:Dmin}). {\bf Left panel:} NFW model, $x$ axis is $\kappa_s$. {\bf Right panel:} NSIS model, $x$ axis is $x_c$.}
     \label{Dmin}
     \end{figure}

We test the validity of the pseudo elliptical approximation by fitting an ellipse  to the isodensity contours of $\kappa_\epsilon$ (obtained making $\kappa_\epsilon=\kappa_{\rm const}$ in equation~\ref{eq:kape}) at the intersection of the tangential critical curve with the $x$ axis ($x_t|_{\phi=0}$; $x$ is the major axis). 
The fits were performed using the $\chi^2$ minimization technique assuming the same error to a discretization of $N$ angularly equidistant points ($\phi_{i+1}-\phi_i=\Delta \phi$). For this section the fits were performed in one quadrant ($\phi_1=0$ and $\phi_N=\pi/2$) and a value of  $N=1000$ was used. 

As in \cite{2012A&A...544A..83D} we measure the goodness of the fit by quantifying the difference of the radial coordinate of the contour $r_\epsilon(\phi_i)$ and the radial coordinate of a fitted ellipse $r_\Sigma(\phi_i)=(a_\Sigma b_\Sigma)/\sqrt{(b_\Sigma {\rm cos}\phi)^2+(a_\Sigma{\rm sin}\phi)^2}$ (with $a_\Sigma$ and $b_\Sigma$ semi major and semi minor axis) from:
\begin{equation} \label{eq:Dmin}
\mathcal{D}^2=\frac{\sum_{i=1}^{N}[r_{\epsilon}(\phi_i)-r_{\Sigma}(\phi_i)]^2}{\sum_{i=1}^{N} r_{\epsilon}(\phi_i)^2}.
	\end{equation}
A value of $\mathcal{D}^2$ less than $\mathcal{D}^2_{\rm min}=4.5\cdot10^{-4}$ indicates an acceptable minimum chi-squared fit ($\chi_\nu^2 < \nu$; $\nu=N-2$) with errors approximately equal to a factor $\mathcal{D}_{\rm min}$ of the squared root mean of the radial points.\footnote{Defining $\chi_\nu^2=(\nu\,\mathcal{D}^2)/\mathcal{D}^2_{\rm min}=\sum_{i=1}^{N}[r_{\epsilon}(\phi_i)-r_{\Sigma}(\phi_i)]^2/\sigma^2$ where $\nu=N-2$. 
Under the assumption that the elliptical model is representative of the data and the errors $\sigma=\mathcal{D}_{\rm min} \sqrt{\sum_{i=1}^N r_\epsilon^2/\nu}$ are product of random fluctuations,  we expect that $\chi_\nu^2$ should follow a chi squared distribution with $\nu$ degrees of freedom with mean equal to $\nu$ and variance equal to $2\nu$.} 
We show comparisons of the pseudo-elliptical ellipticity $\epsilon$ with the fitted ellipticity $\epsilon_\Sigma=1-b_\Sigma/a_\Sigma$ for NFW models in Figure~\ref{epmxnfw} and NSIS models in Figure~\ref{epmxnsis}.  
In Figure~\ref{Dmin}, left panel for NFW models and right panel for NSIS models,  we also show the maximum values $\epsilon$ ($\mathcal{D}^2=\mathcal{D}^2_{\rm min}$) at which the pseudo elliptical approximation remains valid.
From Figures~\ref{epmxnfw}, \ref{epmxnsis} and \ref{Dmin} we find that for both the NFW and NSIS models the pseudo elliptical approximation in general remains valid for $\epsilon \lesssim 0.3$ (or $\epsilon_\Sigma \lesssim 0.5$). 
However, for some extreme cases, like big values of $\kappa_s > 1.5$ in the NFW profile, or models with extreme low values of $x_c$ in the NSIS profile, the pseudo elliptical approximation breaks down at $\epsilon \sim 0.2$.

\section{Survey of Abell clusters with strong lensing} \label{S:lenc}

\begin{figure}
   \includegraphics[width=8.4cm]{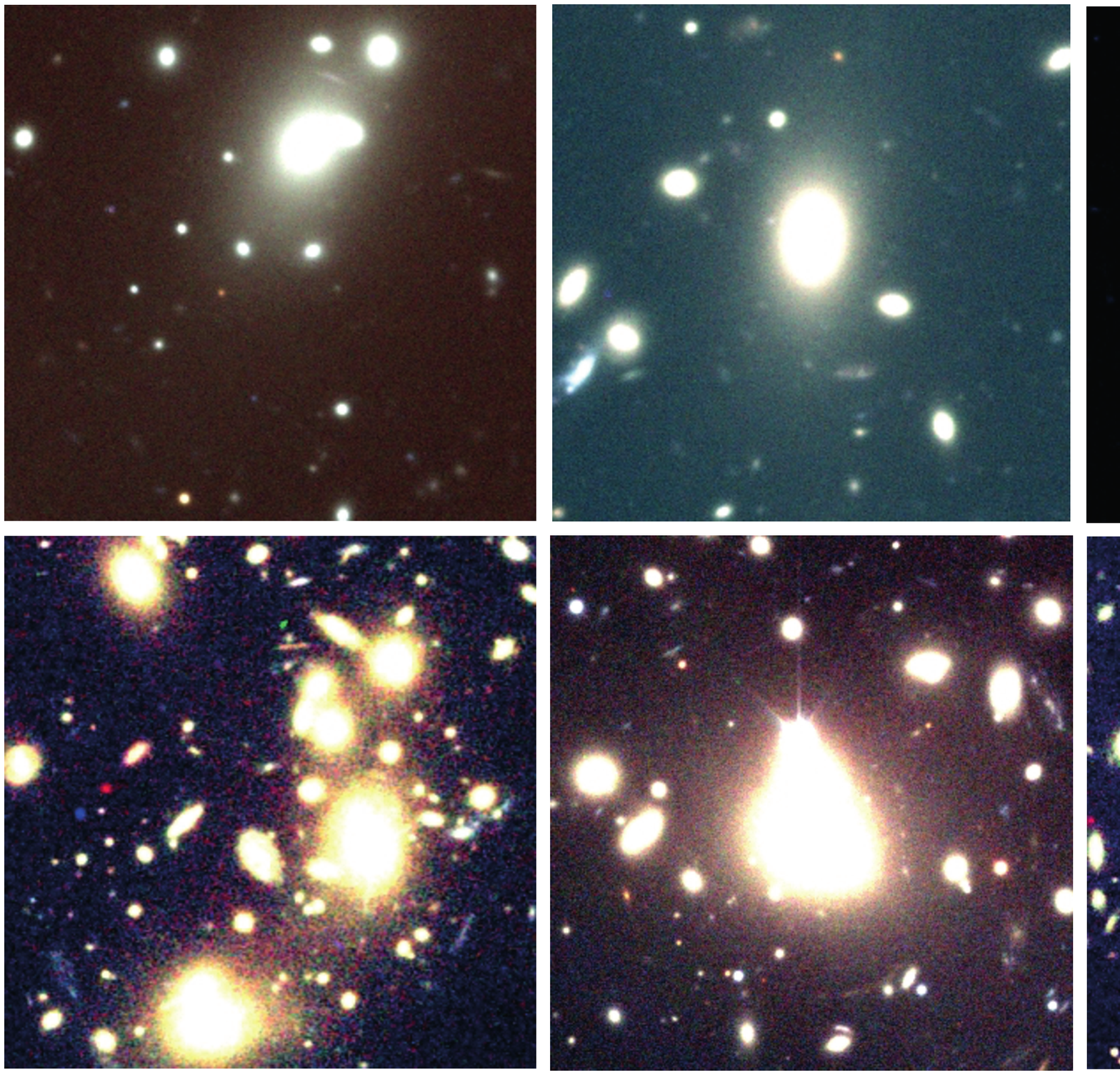}
        \centering
     \caption{Strong lensing candidates in out X-ray bright Abell Cluster sample. Each panel is a $60\arcsec\!\times60\arcsec$ region around the BCG; north is up and East to the left. The panels starting from top-left clockwise are: A0907, A1084, A1451, A2029, A2104, A2163, A2204 and A2744.}
     \label{fig:Lens}
     \end{figure}

In our sample of Abell Clusters, we looked for strong lensing by searching for tangentially elongated structures in the proximities (within 1\arcmin) of the BCG. 
We found evidence of these structures in 8 (out of 49) clusters,  A0907, A1084, A1451, A2029, A2104, A2163, A2204 and A2744 (see Figure~\ref{fig:Lens}).
In the rest of this section, we provide a brief description of the lensing structures (see Figure~\ref{fig:Lens}) and references for cases (5 out of eight) where strong lensing  have been already found in the literature. 

{\bf A0907:} There is one elongated arc in the North west side of the BCG.

{\bf A1084:} Evidence of strong lensing was previously noticed in  \cite{2005ApJ...627...32S}.  
As shown in Figure~\ref{fig:Lens}, there are two arc like structures that are tracing a circular region around the BCG of $\approx19\arcsec$ of radius. The brightest arc like structure is at the southern side of the BCG, the other arc like structure is at the North western side of the BGC. Given the distribution of the structures it is very likely that these are images of the same galaxy.

{\bf A1451:} There is one significantly bright arc with length $\approx 10\arcsec$ at the Northern side of the BCG. 

{\bf A2029:} There is one $\approx 16\arcsec$ elongated arc it the Southern tip of the BCG that looks very faint in Figure~\ref{fig:Lens} due to saturation of light from the central galaxy. This is the lowest redshift cluster in our sample ($z=0.077$) with evidence of strong lensing.

{\bf A2104:} There is one bright $\approx12\arcsec$ arc it the North-eastern tip of the BCG as already noticed by \cite{1994A&A...289L..37P}.

{\bf A2163:} As noticed by \cite{1995ApJ...449...18M}, there are two tangentially elongated galaxies in the south-western side of the BCG. These two red color galaxies in Figure~\ref{fig:Lens} form an arclike structure of $\approx18\arcsec$.

{\bf A2204:} Evidence of strong lensing was previously noticed in  \cite{2005ApJ...627...32S} and \cite{2010MNRAS.404..325R}. As shown in Figure~\ref{fig:Lens}, there are several conspicuous arclets forming a circular structure around the central BGG. 
The most prominent strong lensing feature is a  $\approx 15\arcsec$ arc at the south of the BCG. 

{\bf A2744:}  Evidence of arcs and detailed strong lensing models of this cluster are presented in \cite{2005ApJ...627...32S} and \cite{2014ApJ...797...48J} respectively. 
In our observations, as shown in Figure~\ref{fig:Lens} the most clear arc candidate shows in blue color at the south-western side of the BCG. Notice that surrounding the BCG there other two smaller blue arclets as well.
%

\label{lastpage}

\end{document}